\documentclass[10pt,journal,compsoc]{IEEEtran}
\IEEEoverridecommandlockouts

\usepackage{xcolor}
\usepackage{graphicx}
\usepackage{amsmath}
\usepackage{threeparttable}
\usepackage{enumitem}
\usepackage[newfloat,frozencache,cachedir=.]{minted}
\usepackage{multirow}
\usepackage{siunitx}
\usepackage{tabulary}
\usepackage{booktabs}
\usepackage{comment}
\usepackage[nocompress]{cite}
\usepackage{tikz}
\usepackage{textcomp}
\usepackage{hyperref}

\newcommand\copyrighttext{%
  \scriptsize This is a post peer-review accepted manuscript; published version available online at \href{https://ieeexplore.ieee.org/document/9381618}{ieeexplore.ieee.org/document/9381618} (doi: 10.1109/TC.2021.3066883). \\
  \textcopyright 2021 IEEE. Personal use of this material is permitted.
  Permission from IEEE must be obtained for all other uses, in any current or future
  media, including reprinting/republishing this material for advertising or promotional
  purposes, creating new collective works, for resale or redistribution to servers or
  lists, or reuse of any copyrighted component of this work in other works.}
\newcommand\copyrightnotice{%
\begin{tikzpicture}[remember picture,overlay]
\node[anchor=south,yshift=3pt] at (current page.south) {\fbox{\parbox{\dimexpr\textwidth-\fboxsep-\fboxrule\relax}{\copyrighttext}}};
\end{tikzpicture}%
}

\newcommand{\rev}[1]{\textcolor{black}{#1}}
\newcommand{\revb}[1]{\textcolor{black}{#1}}
\newcommand{\fcrev}[1]{\textcolor{black}{#1}}
\newcommand{\revTCOMP}[1]{\textcolor{black}{#1}}
\newlist{shortletterenum}{enumerate}{10}
\setlist[shortletterenum]{label*=\alph*.,nosep,leftmargin=*}
\newlist{shortenum}{enumerate}{10}
\setlist[shortenum]{label*=\arabic*.,nosep,leftmargin=*}
\newlist{shortitem}{itemize}{10}
\setlist[shortitem]{label*=-,nosep,leftmargin=*}

\definecolor{darkspringgreen}{rgb}{0.09, 0.45, 0.27}

\begin{document}
\title{DORY: Automatic End-to-End\\ Deployment of Real-World DNNs\\ on Low-Cost IoT MCUs}
\vspace{-0.6cm}

\author{Alessio~Burrello,
        Angelo~Garofalo,
        Nazareno~Bruschi,\\
        Giuseppe~Tagliavini,~\IEEEmembership{Member,~IEEE},
        Davide~Rossi,
        Francesco~Conti,~\IEEEmembership{Member,~IEEE}
\IEEEcompsocitemizethanks{\IEEEcompsocthanksitem A. Burrello, A. Garofalo, N. Bruschi, D. Rossi and F. Conti are with the Department of Electrical, Electronic and Information Engineering, University of Bologna, 40136 Bologna, Italy.\protect\\
G. Tagliavini is with the Department of Computer Science and Engineering, University of Bologna, 40136 Bologna, Italy.\protect\\
E-mail: alessio.burrello@unibo.it,  angelo.garofalo@unibo.it,  nazareno.bruschi@unibo.it,  giuseppe.tagliavini@unibo.it,  davide.rossi@unibo.it, f.conti@unibo.it
\IEEEcompsocthanksitem This work was supported in part by the EU Horizon 2020 Research and Innovation projects OPRECOMP  (Open   trans-PREcision COMPuting, g.a. no. 732631) and WiPLASH (Wireless  Plasticity  for  Heterogeneous  Massive  Computer  Architectures, g.a. no. 863337) and by the ECSEL Horizon 2020 project AI4DI (Artificial Intelligence for Digital Industry, g.a. no. 826060).
}
\thanks{This work has been submitted to the IEEE for possible publication. Copyright may be transferred without notice, after which this version may no longer be accessible.}}



\IEEEtitleabstractindextext{
\vspace{-0.6cm}
\begin{abstract}
The deployment of Deep Neural Networks (DNNs) on end-nodes at the extreme edge of the Internet-of-Things is a critical enabler to support pervasive Deep Learning-enhanced applications.
Low-Cost MCU-based end-nodes have limited on-chip memory and often replace caches with scratchpads, to reduce area overheads and increase energy efficiency -- requiring explicit DMA-based memory transfers between different levels of the memory hierarchy.
Mapping modern DNNs on these systems requires aggressive topology-dependent tiling and double-buffering. 
In this work, we propose \textit{DORY} (\textit{Deployment Oriented to memoRY}) -- an automatic tool to deploy DNNs on low cost MCUs with typically less than 1MB of on-chip SRAM memory. 
DORY abstracts tiling as a Constraint Programming~(CP) problem: it maximizes L1 memory utilization under the topological constraints imposed by each DNN layer.
Then, it generates ANSI C code to orchestrate off- and on-chip transfers and computation phases. %
Furthermore, to maximize speed, DORY augments the CP formulation with heuristics promoting performance-effective tile sizes.
As a case study for DORY, we target GreenWaves Technologies GAP8, one of the most advanced parallel ultra-low power MCU-class devices on the market.
On this device, DORY achieves up to 2.5$\times$ better MAC/cycle than the GreenWaves proprietary software solution and 18.1$\times$ better than the state-of-the-art result on an STM32-H743 MCU on single layers.
Using our tool, GAP-8 can perform end-to-end inference of a \textit{1.0-MobileNet-128} network consuming just 63 pJ/MAC on average @ 4.3 fps -- 15.4$\times$ better than an  STM32-H743.
We release all our developments -- the DORY framework, the optimized backend kernels, and the related heuristics -- as open-source software.
\end{abstract}

}
\maketitle
\copyrightnotice
\IEEEdisplaynontitleabstractindextext
\IEEEraisesectionheading{\section{Introduction}}
\IEEEPARstart{T}{he} Internet of Things (IoT) envisions billions of wireless-connected end-nodes~\cite{MAHDAVINEJAD2018161}, which can sense, process and transmit data for a wide range of applications such as surveillance~\cite{motlagh2017uav}, health monitoring~\cite{zanghieri2019robust}, agriculture~\cite{elijah2018overview}, robotics~\cite{palossi201964mw}, and others.
However, major challenges are linked to this new computation paradigm, including reliability, security, capacity, together with the production of high-bandwidth data.
In this scenario, edge-based Deep Learning (DL) is an attractive approach thanks to its capability to extract high-level features from raw sensor data, reducing off-node transmissions, and improving security by doing most processing in-place.

Modern Deep Neural Network (DNN) inference tasks run on cloud servers, personal computers, or smartphones. Even in the most constrained scenario of mobile devices, their execution can count on GB of memory and significant processing power available, under a power envelope of a few watts.
Conversely, deploying DNNs on a microcontroller-based IoT end-node has to deliver similar performance while dealing with \textit{i})  strict constraints in terms of memory (a few MB off-chip, and typically 1 MB on-chip at most), \textit{ii}) limited computational capabilities, and \textit{iii}) battery constraints and a peak power envelope of 100-200 mW.
The deployment of DL-based algorithms on the IoT demands aggressive hardware, software, and algorithmic co-optimization to exploit the scarce resources on these systems to the maximum degree~\cite{conti2017iot}.
In particular, the scarce availability of memory constitutes a real Deep Learning Memory Wall~\cite{conti2020memory}: a fundamental limitation to the maximum performance of an embedded DNN compute system.

Recently introduced algorithmic improvements such as quantized DNN  inference~\cite{gao2018ifq} aim at matching a DNN's full-precision accuracy while using exclusively 8-bit (or smaller) integer data to reduce memory occupation and execution complexity. On the hardware side, accelerators~\cite{7551407,7092475,sze2017hardware} and instruction set architecture (ISA) extensions~\cite{garofalo2020xpulpnn} that exploit quantization have been introduced to speed up the computation, lessen the impact of memory constraints and minimize energy consumption. In essence, 8-bit networks are now supported by most of the frameworks, such as TensorFlow and PyTorch.
Recently proposed architectural paradigms aim at maximizing DNN performance and efficiency on IoT end-nodes while safeguarding the flexibility of typical Microcontroller Unit (MCUs), so that common control-oriented MCU tasks can be mixed with DNNs and non-DL-based data processing tasks.
These architectures often couple a conventional MCU with an accelerator~\cite{7870349,garofalo2020pulp}.
\textit{Parallel Ultra-Low-Power computing} (PULP), for example, is an architectural paradigm based on flexible software-oriented acceleration for DNNs and other data processing tasks in multi-core end-nodes.
The core idea of PULP is to couple an I/O-dedicated core with a multi-core cluster of processors optimized for data-parallel processing, sharing a high-bandwidth multi-banked L1 memory~\cite{conti2016pulp}.

Accelerated IoT end-nodes employ multi-level hierarchies of on- and off-chip memories. In some cases, they
do away entirely with energy-expensive coherent data caches, exploiting manually managed scratchpad memories instead to maximize area and energy efficiency.
For example, PULP architectures complement a small ($<$ 128 kB) L1 with a bigger-but-slower ($\sim$1 GB/s) on-chip L2 memory, and by an off-chip L3 low-power IoT DRAM~\cite{hyperram} that provides high capacity, but at a slower speed ($\sim$100 MB/s) and with relatively high energy penalty ($>$ 50 pJ/B).
These composite memory hierarchies are becoming necessary even in low-power systems to cope with the memory footprint of DNN inference, without paying the cost of huge on-chip caches.
However, to  ``unlock'' such a system's theoretical performance often requires carefully managed data movement by means of cache locking or explicit DMA transfers.
%
To reduce the related development overhead, software caches~\cite{saidi2013optimizing} and data tiling strategies~\cite{tagliavini2015adrenaline} have been proposed: however, most DL-based applications can improve upon general-purpose solutions by exploiting the regular structure of DNNs, with ad-hoc memory management flows to minimize inference time~\cite{palossi201964mw,cecconi2017optimal}, exploit data reuse, and optimize scheduling~\cite{peemen2013memory}.
Conversely, automatic solutions for end-to-end deployment of real-world DNNs on MCUs so-far rely either on slow and inefficient interpretation (e.g., TF-Lite Micro~\cite{david2020tensorflow}), or on proprietary code generation frameworks (e.g., ST XCUBE-AI~\cite{CubeAI}, GWT\footnote{GreenWaves Technologies.} AutoTiler~\footnote{\texttt{\url{https://greenwaves-technologies.com/manuals/BUILD/AUTOTILER/html/index.html}}}).

In this paper, we introduce a novel lightweight framework called DORY, Development Oriented to memoRY, which aims at the deployment of end-to-end DNNs on memory-starved end-nodes, and particularly tuned to the class of end-nodes based on the PULP paradigm.
As our main case study, we target GWT GAP-8~\cite{flamand2018gap} -- one of the most advanced low-power edge nodes available in the market, embodying the PULP architectural paradigm with DSP-enhanced RISC-V cores.
%



We introduce several novel contributions:
\begin{shortenum}
    \item \revTCOMP{A tool for multi-level memory tiling aiming at the deployment of realistically sized DNNs on memory-starved MCUs. Relying on Constraint Programming (CP) optimization, our tool matches on- and off-chip memory hierarchy constraints with DNN geometrical requirements, such as the relationships between input, weight, and output tensor dimensions.}
    \item A set of heuristics to maximize the performance of the CP solution on PULP platforms using the dedicated backend library PULP-NN~\cite{garofalo2020pulp}, to maximize throughput and energy efficiency in the RISC-V based GAP-8 target. 
    \item \revTCOMP{A code generator using tiling solutions to produce ANSI C code for the target platform, with all data L3-L2-L1 orchestration implemented as fully pipelined, triple-buffered DMA transfers and integrated calls to the computational backend (PULP-NN).}
    \item \revTCOMP{Tailored optimizations for common architectural features of modern DNNs: i) for residual connections, a bidirectional stack avoiding memory fragmentation; ii) for depthwise layers, a new optimized backend not present in PULP-NN.}
\end{shortenum}
We evaluate the performance and energy efficiency of the deployed networks produced by DORY on GWT GAP-8, considering both single layers and end-to-end networks.
DORY achieves up to 18.1$\times$ better MAC/cycle than the state-of-the-art result on a conventional cache-based microcontroller, the STM32-H743 MCU, in single layer execution. 
Using DORY, end-to-end deployment of 8-bit quantized networks such as \textit{0.5-MobileNet-v1-192}, achieve up to 8.00 MACs/cycle, with a 13.2$\times$ improvement compared to the same networks running on the STM32-H743 using the state-of-the-art ST X-CUBE-AI. Furthermore, on a layer by layer basis, DORY can achieve up to 2.5$\times$ better throughput than the proprietary GWT AutoTiler, on the same GAP-8 platform, and up to 27\% better performance on full network execution.
Our results show that image recognition on an extreme edge-node can run in as little as 11.9 mJ/classification @ 4.3 fps.

\revTCOMP{To put our results into context, we compare the efficacy of DORY on GAP-8 with that obtainable in a state-of-the-art single-core ARM M7 core, the off-the-shelf ST32-H743 MCU with 16 kB of L1 data cache (D\$) and 128 kB of L1 scratchpad memory.
For a set of 44 DNN layers of different size, we compare i) single-core execution on GAP-8 with DORY-based memory management, ii) M7 execution with active D\$, iii) M7 execution with DORY-managed DMA transfers on the scratchpad memory.
Our results show that on the M7, DORY automatic memory management is up to 9\% faster than the 16 kB hardware cache, and never slower.
We also show that single-core execution on GAP-8 is, on average, 2.5$\times$ faster than on the M7 in cycles/cycles thanks to the full exploitation of the DSP-enhanced RISC-V cores.
}

To foster research on real-world deeply embedded DNN applications, we release the DORY framework, the optimized backend kernels, and the PULP heuristics discussed in this paper as open-source \footnote{{\texttt{\href{https://github.com/pulp-platform/dory}{https://github.com/pulp-platform/dory}}}}.

\section{Related Work}
\label{sec:related}

\subsubsection*{DNN algorithm minimization}

\begin{table*}
\footnotesize
\centering
{
\centering
{\color{black}
\caption{Data flow scheduling and tiling in literature for different computing scales, super computing, ASIC accelerators, and tiny MCUs.\label{tab:SoA_Related}}
\vspace{-0.3cm}
\begin{tabular}{p{1.7cm}p{1.7cm}llp{2cm}l}
\multicolumn{1}{l|}{\textbf{Work}}         & \textbf{Networks}                                                       & \textbf{Optimizations}                                                                                          & \textbf{Output}                                                                                   & \textbf{Open-Source} & \textbf{Precision}  \\ \hline \hline
\multicolumn{6}{l}{\textbf{Supercomputers}}                                                                                                                                                                                                                                                                                       \\ \hline  \hline
\multicolumn{1}{l|}{DMIAYN~\cite{ivanov2020data}}       & Transformers                                                   & \begin{tabular}[c]{@{}l@{}}1) Operator Fusing, \\ 2) Data Layout Exploration\end{tabular}              & Transformer Primitives                                                                   & Yes         & fp32       \\ \hline \hline
\multicolumn{6}{l}{\textbf{DNN   Accelerators}}                                                                                                                                                                                                                                                                                   \\ \hline \hline
\multicolumn{1}{l|}{dMazeRunner~\cite{dave2019dmazerunner}}  & \begin{tabular}[c]{@{}l@{}}CNN,   \\ Nested Loops\end{tabular} & \begin{tabular}[c]{@{}l@{}}1) Loop   Ordering, \\ 2) Loop Tiling, \\ 3) Memory Movements\end{tabular}  & \begin{tabular}[c]{@{}l@{}}Temporal/Spatial Schedule, \\ Loop Tiling\end{tabular}      & Yes         & Flexible   \\ \hline
\multicolumn{1}{l|}{MAESTRO~\cite{kwon2020maestro}}      & CNN                                                            & \begin{tabular}[c]{@{}l@{}}1) Mapping \& Data Reuse, \\ 2) PEs Design\end{tabular}                     & \begin{tabular}[c]{@{}l@{}}PEs array,\\ Temporal/Spatial Schedule\end{tabular}         & Yes         & Flexible   \\ \hline
\multicolumn{1}{l|}{Interstellar~\cite{yang2020interstellar}} & \begin{tabular}[c]{@{}l@{}}CNN,   \\ LSTM, \\ MLP\end{tabular} & \begin{tabular}[c]{@{}l@{}}1) Loop Ordering, \\ 2) Loop Tiling, \\ 3) PEs+Mem. Design\end{tabular}     & \begin{tabular}[c]{@{}l@{}}PEs + Mem. Array, \\ 7-Loops Ordering and Tiling\end{tabular} & Yes         & 16   bits  \\ \hline
\multicolumn{1}{l|}{Timeloop~\cite{parashar2019timeloop}}     & CNN                                                            & \begin{tabular}[c]{@{}l@{}}1) Loop Ordering\\ 2) Loop Tiling\end{tabular}                              & \begin{tabular}[c]{@{}l@{}}Model Scheduling, \\ Latency/Energy Estimation\end{tabular}   & No          & Flexible   \\ \hline \hline
\multicolumn{6}{l}{\textbf{Mobile   \& MCUs}} \\ \hline \hline
\multicolumn{1}{l|}{LCE~\cite{Larq}}          & BNN                                                            & \begin{tabular}[c]{@{}l@{}}1) Loop Tiling, \\ 2) Vectorization, \\ 3) Parallelization\end{tabular}     & \begin{tabular}[c]{@{}l@{}}C++ Runtime Interpreter,\\ C++ Descriptor\end{tabular}        & Yes         & 1bit       \\ \hline
\multicolumn{1}{l|}{TFLite Micro~\cite{david2020tensorflow}}       & \begin{tabular}[c]{@{}l@{}}CNN,\\ MLP\end{tabular}             & \begin{tabular}[c]{@{}l@{}}1) Hand-configurable Mem., \\ 2) Optimized Backends\end{tabular}            & \begin{tabular}[c]{@{}l@{}}C++ Runtime Interpreter,\\ C++ Descriptor\end{tabular}        & Yes         & int8-fp32  \\ \hline
\multicolumn{1}{l|}{Cube-AI~\cite{CubeAI}}      & \begin{tabular}[c]{@{}l@{}}CNN,   \\ MLP\end{tabular}          & 1) Mem. Access Opt.                                                                                    & C Optimized Executable                                                                   & No          & int8-fp32  \\ \hline
\multicolumn{1}{l|}{GWT AutoTiler} & \begin{tabular}[c]{@{}l@{}}CNN,   \\ MLP\end{tabular}          & \begin{tabular}[c]{@{}l@{}}1) Loop Tiling, \\ 2) Mem. Access Opt.\end{tabular}                         & C Optimized Executable                                                                   & Partially   & int8-int16 \\ \hline
\multicolumn{1}{l|}{DORY}         & \begin{tabular}[c]{@{}l@{}}CNN,   \\ MLP\end{tabular}          & \begin{tabular}[c]{@{}l@{}}1) Loop Tiling, \\ 2) Mem. Access Opt.\\ 3) Mem. Fragmentation\end{tabular} & C Optimized Executable                                                                   & Yes         & int8       \\ \hline
\end{tabular}}
}
\vspace{-0.4cm}
\end{table*}

From the algorithmic viewpoint, the first task in DL deployment is making sure that the DNNs are ``minimally redundant'', in the sense that they do not perform any additional operation unless it leads to a better quality-of-results.
In this direction, a current research trend is to adapt DNN architectures to deployment in constrained platforms by shrinking the DNN topologies themselves, either directly~\cite{howard2017mobilenets,s2018mobilenetv2} or using neural architecture search~\cite{tan2019efficientnet, lin2020mcunet}.
Orthogonally, system designers can adopt techniques for post-training quantization~\cite{capotondi2020cmix} and quantization-aware fine-tuning~\cite{choi2018pact} to reduce the cost of single operations in terms of energy and of single parameters in terms of memory -- trying to minimize the price in terms of quality-of-results.

\subsubsection*{Optimized software \& ISA for DNN computation}
Given a size-optimized and precision-tuned DNN, we need to address the deployment challenge, i.e., achieve maximal utilization of the computing units, while minimizing the performance and energy penalties associated with data transfers across the memory hierarchy. 
Application-specific hardware architectures are very useful in accelerating particular layers and, in some cases, entire networks~\cite{7551407,7092475, sze2017hardware} -- but their lack of flexibility can be a liability in a field such as DL, where every year researchers introduce tens of new topologies and different ways to combine the DNN basic blocks.
To provide higher flexibility, in many cases, DNN primitives are implemented in highly optimized software instead of full-hardware blocks.
Over the years, several software libraries of DNN kernels have been proposed~\cite{garofalo2020pulp, lai2018cmsis, capotondi2020cmix, rusci2018work} to maximize the efficiency of DNN execution with DSP-oriented single-instruction multiple-data (SIMD) ISA capabilities~\cite{gautschi2017near}.
These libraries leverage either the Height-Width-Channel (HWC) or Channel-Height-Width (CHW) data layout to minimize operations and memory footprint.
CHW optimizes data reuse in the spatial dimensions. Therefore, it is faster on convolutions with larger filters and lower channel connectivity; HWC naturally favors channel-wise data reuse, often requiring the construction of a flattened data structure ('im2col' buffer) to exploit spatial data reuse partially~\cite{lai2018cmsis}.
Further, there is an increasing trend towards more targeted ISA specialization (e.g., ARM Helium~\footnote{\texttt{\url{https://www.arm.com/why-arm/technologies/helium}}}, xPULPNN~\cite{garofalo2020xpulpnn}) to support and accelerate the pervasive convolutional layers with low-bitwidth linear algebra instructions.

\revTCOMP{
\subsubsection*{Memory hierarchy management}
One of the most critical challenges in DNN deployment is memory hierarchy management: modern DNNs generate high amounts of weight and activation traffic between different levels of the memory hierarchy, which may constitute a significant bottleneck. 
In Table~\ref{tab:SoA_Related}, we report different methods for data flow scheduling and generation that cover three broad classes of devices, namely high-performance computing systems~\cite{ivanov2020data}, DNN accelerators~\cite{dave2019dmazerunner,kwon2020maestro,yang2020interstellar,parashar2019timeloop}, and embedded systems~\cite{tf-micro,Larq}.
For what concerns high-performance computing systems, \cite{ivanov2020data} propose new transformer primitives to exploit data reuse and limit data movement by fusing pointwise operators.
On the other hand, \cite{dave2019dmazerunner,kwon2020maestro,yang2020interstellar,parashar2019timeloop} discuss DNN optimization on AI-specialized accelerators based on systolic arrays of processing elements (PEs), with a focus on loop tiling and/or reordering to i) efficiently move the data to fastest memory regions and ii) correctly schedule layers in space and time to maximize PE utilization.
The output of these tools can be either an accelerator model to run a given DNN~\cite{kwon2020maestro,yang2020interstellar} or the spatial scheduling to maximize PE array utilization on a target accelerator \cite{dave2019dmazerunner, parashar2019timeloop}.}

\revTCOMP{
MCU data flow scheduling tools show similarities to frameworks such as DMazeRunner, as both target the optimization of a dataflow schedule given an externally known architecture. However, the MCU scenario also imposes some additional unique challenges, such as the fact that DNN execution has to be adapted to a general-purpose architecture and the small amount of memory that MCU platforms include. 
Further, the kernel instructions are heavily influenced by the limited size of the register file, which causes additional load-store operations and thus demand for an optimal loop sizing to avoid register spilling overhead.
Academic researchers and industries have significantly investigated this aspect by including in their edge-node solutions either specialized caches (e.g., NXP \footnote{\texttt{\url{https://www.nxp.com/products/processors-and-microcontrollers/arm-microcontrollers/general-purpose-mcus/lpc4300-cortex-m4-m0}}}) or explicitly managed scratchpad memories (e.g., GWT \cite{flamand2018gap}).}
%
%
%
%
%
%

\subsubsection*{DNN-oriented microcontrollers and related tools}
Recently, the first generation of low-power neural-network oriented MCUs has been introduced,
coupling optimized software and ISA extensions for DNN computing with ``traditional'' control and I/O-bound activities.
To enable optimal execution of both kinds of tasks, these MCUs exploit parallel and heterogeneous processing; for example, ST Microelectronics\footnote{\texttt{\url{https://www.st.com/en/microcontrollers-microprocessors/stm32h7-series.html}}} and NXP have recently introduced new-generation dual-core microcontrollers with an ARM M0 processor dedicated to I/O and an ARM M4 processor with single-cycle multiply-and-accumulate and SIMD capabilities.
These platforms show an increased complexity in terms of memory hierarchy compared to conventional flat-memory MCUs, with an L1 memory optimized for speed and an L2 optimized for capacity.
At the same time, there is a trend towards explicit management of memory hierarchy, with hand-tunable data caches featuring locking for hand-crafted data management.
To manage this complexity, these MCUs include dedicated infrastructure for data marshaling, such as general-purpose DMA controllers to speed-up memory transfers and reduce the memory access bottleneck.

\revTCOMP{New platforms magnify these industry-wide architectural trends, introducing multi-core and AI-specific accelerators and removing data caches, replacing them with small on-chip scratchpad memories.
For instance, the Kendrite K210~\footnote{\texttt{https://canaan.io/product/kendryteai}} is a RISC-V dual-core 64 bits system-on-chip with a neural network processor (KPU) on which the cores can offload the computation. It also includes dedicated memory banks for the NN accelerator and a DMA unit to explicitly manage the transfers.
The SONY Spresense board~\footnote{\texttt{https://developer.sony.com/develop/spresense/}} features a 6-cores M4 accelerator with a maximum clock speed of 156 MHz, 1.5 MB of SRAM and 8 MB of Flash.
The GreenWaves Technologies GAP-8~\cite{flamand2018gap} system-on-chip, which we target as a case study in this work, was introduced in 2018 as a commercial embodiment of the \textit{Parallel Ultra-Low-Power} paradigm~\cite{conti2016pulp}: it features one I/O core and an 8-core SIMD-optimized DSP cluster accelerator using an extension of the RISC-V ISA.}
\revTCOMP{
Programming these DNN-oriented MCUs is typically more complicated with respect to conventional MCUs.
Maximizing the exploitation of computational resources is challenging, and scratchpads require manually managed data orchestration and tiling.
}

\revTCOMP{
New tools such as TFLite Micro~\cite{david2020tensorflow} and the Larq Computing Engine (LCE)~\cite{Larq} offer a model-agnostic deployment framework and overcome these problems.
Both are non-vendor-locked tools supporting ARM Cortex-M and RISC-V cores.
Their library memory footprints require only 16 kB on a Cortex-M3; however, by default they rely on graph interpretation at runtime, limiting achievable performance.
To offset this limitation, TFLite Micro allows plugging in optimized kernels and declaring vectors in different memory regions. However, it does not include any tiling mechanism to execute layers that do not fit on-chip memory.}

\revTCOMP{
To the best of our knowledge, the two most powerful DNN deployment tools available in the state-of-the-art have been proposed by the industry as proprietary, vendor-locked solutions for their own MCUs.
X-CUBE-AI~\cite{CubeAI} from STMicroelectronics is an automatic NN library generator optimized on computation and memory.
It converts a pre-trained DNN model from DNN tools such as Tensorflow into a precompiled library for the ARM Cortex-M cores embedded in STM32 series MCUs.
X-CUBE-AI relies on relatively large on-chip L1 caches (up to 16 kB) to deliver performance on STM32 MCUs, and it does not tackle software-based memory management.
%
On the other hand, GWT designed a tool called AutoTiler, to target the GAP-8 RISC-V based multi-core ultra-low-power microcontroller.
One of its primary functions is to take a pre-trained DNN and generate code for memory tiling and efficient transfers of weight and activation data between all memory levels (on- and off-chip).
The GWT AutoTiler directly tackles the data-movement and tile sizing challenge to optimize memory access, reaching state-of-the-art performance on the execution of many networks.
The tool is proprietary, but its backend basic kernels are available as open-source as part of the GAP-8 SDK\footnote{https://github.com/GreenWaves-Technologies/gap\_sdk}.}

\revTCOMP{
DORY is the first open-source framework to directly tackle the MCU memory hierarchy management challenge, with a comprehensive exploration of data tiling, optimized loop ordering for different layers (i.e., pointwise and depthwise), and a solution for the data fragmentation problem that is critical to deploy residual layers at the edge.
In Section~\ref{sec:results}, we perform several quantitative comparisons with the best results obtained with STM X-CUBE-AI, GWT AutoTiler, and our own DORY. DORY consistently outperforms all the competitors on all the proposed benchmarks.}

\section{Background}
\label{sec:background}

\subsection{Quantized Neural Networks}
\label{sec:qnns}
Post-training quantization~\cite{capotondi2020cmix} or quantization-aware training~\cite{choi2018pact} produce as output a Quantized Neural Network (QNN).  
%
%
%
\revTCOMP{
In the context of this work, we consider QNNs produced with \textit{linear uniform per-layer quantization}, where all tensors $\mathbf{t}$ (e.g., weights $\mathbf{w}$, inputs $\mathbf{x}$, or outputs $\mathbf{y}$) defined in a  range$[\alpha_\mathbf{t}, \beta_\mathbf{t})$ can be mapped to $N$-bit integer tensors $\widehat{\mathbf{t}}$ through a bijective mapping:
\begin{equation}
    \mathbf{t} = \alpha_\mathbf{t} + \varepsilon_\mathbf{t}\cdot \widehat{\mathbf{t}} \label{eq:1}\;, 
\end{equation}
where $\varepsilon_\mathbf{t} = (\beta_\mathbf{t}-\alpha_\mathbf{t}) / (2^{N}-1)$.
We call $\varepsilon_\mathbf{t}$ the \textit{quantum} because it is the smallest amount that we can represent in the quantized tensor.
}

%

Each QNN layer is composed of a sequence of three operators: Linear, Batch-Normalization (optionally) and Quantization/Activation.
Without loss of generality, we consider that $\alpha_\mathbf{x} = \alpha_\mathbf{y} = 0$ for all the inputs of Linear  and the outputs of Quantization/Activation operators\footnote{If the original activation is a ReLU, then the QNN automatically satisfies this condition; otherwise, it can be transformed to satisfy it.}, \revTCOMP{but not for weights}.
%
Using Eq.~\ref{eq:1}, all operators are mapped in the integer domain: 
    \begin{align}
        \mathrm{LIN:} \quad&\varphi = \sum_n \mathbf{w}_{m,n} \mathbf{x}_n \iff \widehat{\varphi} = \sum_n \widehat{\mathbf{w}_{m,n}} \cdot \widehat{\mathbf{x}_n}
        \label{eq:2} \\
        \mathrm{BN\footnotemark:} \quad&{\varphi'} = {\kappa}\cdot {\varphi}+{\lambda} \iff \widehat{\varphi}' = \widehat{\kappa}\cdot \widehat{\varphi}+\widehat{\lambda}
    \label{eq:normalization}
        \;.
    \end{align}
%
\footnotetext{In inference, the statistical and learned parameters of BN can be combined: $\kappa=\gamma/\sigma$ and $\lambda=\beta - \mu\gamma/\sigma$.}
%
The dot product operation in Eq.~\ref{eq:2} results in a shrinking of the quantum used to represent $\widehat{\varphi}$, which will be $\varepsilon_\varphi = \varepsilon_\mathbf{w}\varepsilon_\mathbf{x}$.
Hence, we need to represent the integer output of the Linear operator ($\widehat{\varphi}$) with higher precision (e.g., 32 bits) with respect to its inputs, before re-quantizing it at the end of the accumulation.
A similar consideration applies to Batch-Normalization and its output $\widehat{\varphi}'$.

The final Quantization/Activation operator \textit{i)}~provides a non-linear activation essential for the QNN to work at all, and \textit{ii)}~collapses the accumulator into a smaller bitwidth:
\begin{equation}
    \mathrm{QNT/ACT:} \quad\widehat{\mathbf{y}} = 
    m \cdot \widehat{\varphi}' \gg d \;;\;
    m=\left\lfloor\frac{
        \varepsilon_{\mathbf{\varphi}'}\cdot 2^{d}
    }{
        \varepsilon_\mathbf{y}
    }\right\rfloor\;. 
    \label{eq:requantization}
\end{equation}
$d$ is an integer chosen during the quantization process in such a way that $\varepsilon_\varphi/\varepsilon_\mathbf{y}$ can be represented with sufficient accuracy inside $m$.
A method similar to Eq.~\ref{eq:requantization} is also used when multiple branches of the network, each with its own $\varepsilon$, reconverge in a single tensor (typically using summation).
In that case, the branches are ``matched'' to the same quantum using a variant of Eq.~\ref{eq:requantization}.

Thanks to the mapping of Eq.~\ref{eq:1}, it is possible to execute the entire network using only integer data.
\revTCOMP{In this work, we target networks using 8-bit quantization for both $\widehat{\mathbf{w}}$~(signed), and $\widehat{\mathbf{x}}$ / $\widehat{\mathbf{y}}$~(unsigned); $\widehat{\varphi}$, $\widehat{\varphi}'$, and the $\widehat{\kappa}$, $\widehat{\lambda}$, $m$, $d$ parameters use 32-bit integers (signed).
We relied on the open-source NEMO library~\cite{conti2020nemo} to generate QNN topologies in the format described in this Section.
Note that using different quantization techniques such as non-linear 8 bits quantization or clustering \cite{han2016deep} for network compression and execution would be possible with DORY replacing the software backend employed.}

\subsection{Parallel Ultra-Low-Power computing paradigm}
\label{subsec:pulp}
\begin{figure}
  \centering
\includegraphics[width=0.995\columnwidth]{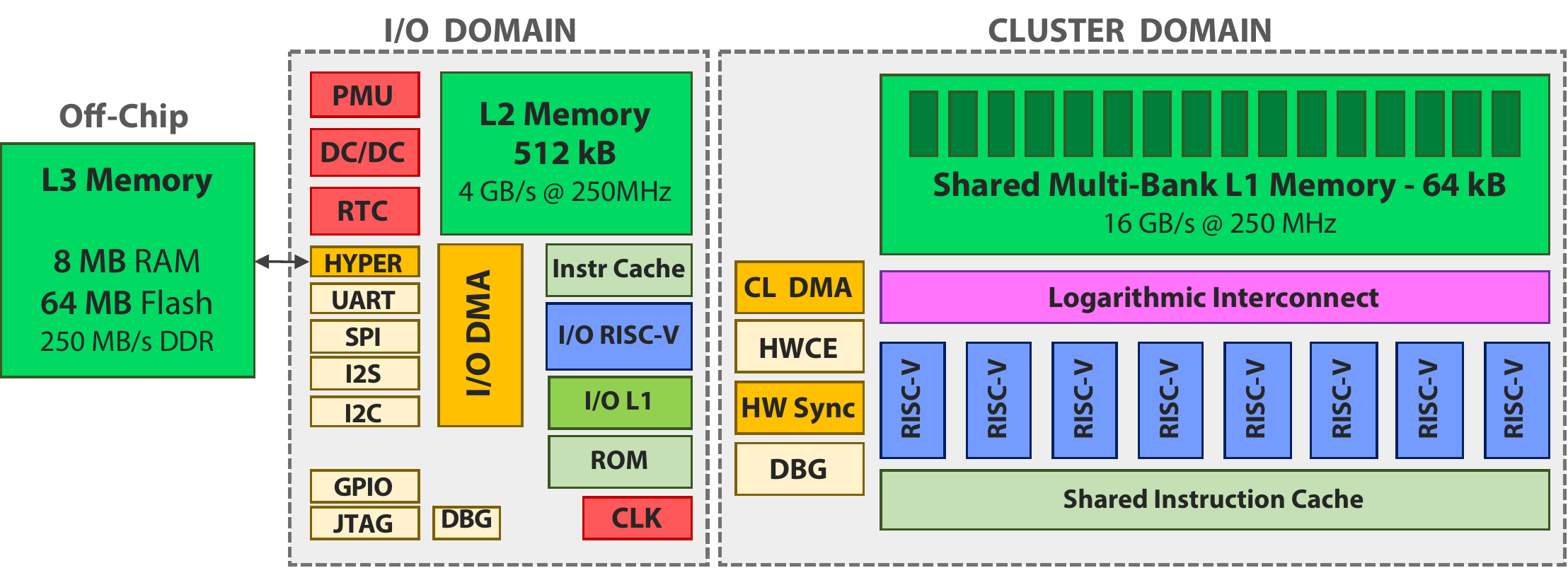}
\vspace{-0.4cm}
  \caption{GWT GAP-8 MCU block diagram.}
\label{fig:GAP8}
\vspace{-0.3cm}
\end{figure}

\begin{figure*}
  \centering
\includegraphics[width=0.65\textwidth]{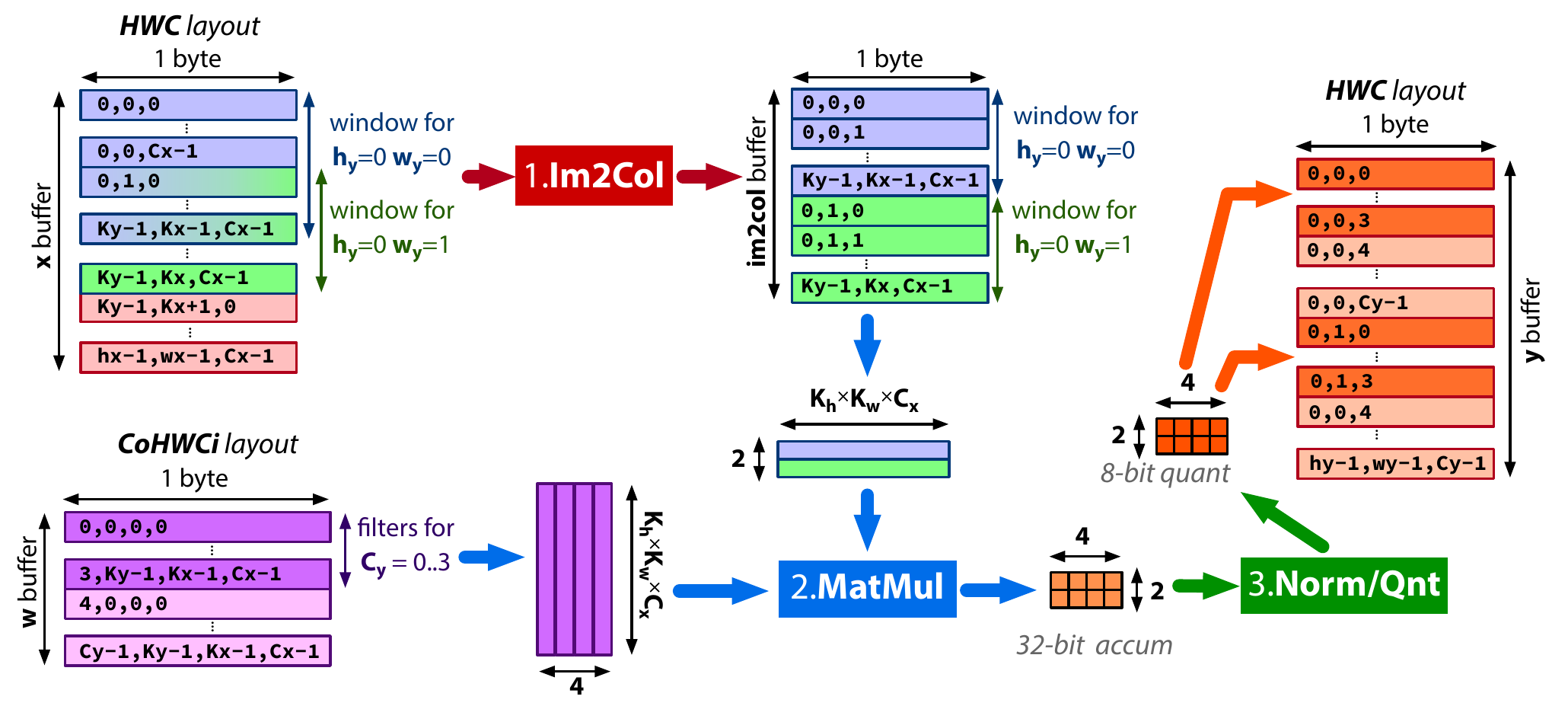}
\vspace{-0.5cm}
  \caption{PULP-NN~\cite{garofalo2020pulp} execution model divided in Im2Col, MatMul, Norm/Qnt phases (see Table~\ref{tab:notation} for the buffer naming scheme).}
\label{fig:pulp_nn_overview}
\vspace{-0.4cm}
\end{figure*}

\revTCOMP{
Research and industry are dedicating increasing attention to edge-nodes with specialized co-processors (accelerators) and hierarchical memories, designed to exploit the pervasive data-regularity in emerging data-analytics tasks (e.g., deep-learning).
}
Parallel Ultra-Low Power computing is an architectural paradigm leveraging near-threshold computing to achieve high energy efficiency, coupled with parallelism to improve the performance degradation at low-voltage~\cite{pullini2019mrwolf}.
The PULP paradigm builds upon the trends explained in Section~\ref{sec:related}: ISA optimizations for DSP and DNN computing; heterogeneous parallel acceleration, with architecturally different compute units dedicated to unrelated tasks; and explicitly managed memory management.
PULP systems are centered around a state-of-the-art single-core microcontroller (\textit{I/O domain}) with a standard set of peripherals.
The I/O core offloads parallel tasks to a software-programmable parallel accelerator composed of N additional cores, standing in its own voltage and frequency domain (\textit{cluster domain}).

GWT GAP-8~\cite{flamand2018gap} (depicted in Figure~\ref{fig:GAP8}) is a commercial PULP system with 9 extended RISC-V cores (one I/O + an eight-core cluster), which we chose as the reference platform in this work since it represents one of the most advanced embodiments of the DNN-dedicated MCU trends.

The GAP-8 'cluster' is composed by eight 4-stage in-order single-issue pipeline RI5CY~\cite{gautschi2017near} cores, implementing the RISC-V RV32IMCXpulpV2 Instruction Set Architecture (ISA). XpulpV2 is a domain-specific extension meant for efficient digital signal processing, with hardware loops, post-modified access LD/ST, and SIMD instructions down to 8-bit vector operands.

The cores of the \textit{cluster} share a first level of memory, a 64 kB multi-banked L1 memory Tightly-Coupled Data Memory (TCDM), accessible from the cluster's cores through a high-bandwidth, single-cycle-latency logarithmic interconnect, featuring a 2$\times$ banking factor and a word-level interleaving scheme to reduce the probability of contention~\cite{rahimi2011fully}. 
In order to manage data transfers between the L1 TCDM memory and a second-level 512 kB of memory (managed as a scratchpad as well) available in the SoC domain, the \textit{cluster DMA}~\cite{rossi2014ultra} can manage data transfers between L1 and L2 with a bandwidth up to 2 GB/s and a latency of 80 ns at the maximum frequency. On the other hand, to interface the L2 memory with the external world, and in particular with the Cypress Semiconductor's HyperRAM/HyperFash module~\cite{hyperram} available on the GAPuino board, GAP-8 can use an autonomous I/O subsystem called \textit{I/O DMA}~\cite{pullini2017mudma}. Through the HyperBus interface, the external L3 HyperRAM and/or HyperFlash memory can be connected to the system, enabling a further 64~MB of storage for read-only data on Flash and 8-16~MB for volatile data on DRAM, with a bandwidth up to 200 MB/s.

\subsection{QNN Execution Model on GAP-8}
\label{subsec:pulpnn}
\begin{table}[t]
\caption{Symbols used throughout this work.}
\vspace{-0.3cm}
\label{tab:notation}
\footnotesize
\centering
\begin{tabular}{lc}
\toprule
Input  $\mathbf{x}$ dims (height/width/chan) & $h_x$ / $w_x$ / $C_x$ \\
Output  $\mathbf{y}$ dims (height/width/chan) & $h_y$ / $w_y$ / $C_y$ \\
Weight $\mathbf{w}$ dims (out c/height/width/in c) & $C_y$ / $K_h$ / $K_w$ / $C_x$ \\ \midrule
Buffer for tensor $q$ at $i$-th level of mem. hier. & $Li_q$ \\
Tiled dimension $d_q$ of a tensor $\mathbf{q}$ & $d_q^t$ \\\bottomrule
\end{tabular}
\vspace{-0.4cm}
\end{table}

Computational backends are by construction tied to a specific target platform as they need to fully exploit the architecture's strength.
As optimized QNN backend for our GAP-8 case study, we relied  on the open-source PULP-NN~\cite{garofalo2020pulp} library.
PULP-NN is based on the HWC data layout.
An efficient QNN layer is implemented in the backend library as a combination of three phases, summarily shown in Figure~\ref{fig:pulp_nn_overview}.
First, the Im2Col step copies the pixels needed to produce a single output pixel (i.e., the \textit{receptive field}) from their 3-D input non-sequential in memory arrangement into a 1-D vector using load/store operations.
Note that this step is not performed for 1$\times$1 convolutions, since all the necessary input pixels ($1 \times 1 \times C_x$) are already sequential in memory, given the HWC data layout.
Then, the linear part of the kernel, the Matrix Multiplication (MatMul), convolves the current 1-D vector with the weight parameters of the layer, exploiting the RI5CY SIMD instructions to implement the integer part of Eq.~\ref{eq:2}.
To improve performance, the innermost loop of the MatMul accumulates the partial results of the convolution over registers, eliminating the store instructions inside the loop and reusing the 1-D input vector elements along with 4 different sets of filters.
This enables the computation of 2 adjacent output pixels in parallel, thus maximizing reuse and reducing the cost of loads.
In this way, the innermost loop consists of just 6 load (\texttt{ld}) instructions and 8 SIMD MAC instructions (\texttt{sdotp}), for a total of 32 MACs per loop iteration.
In this work, we extended the PULP-NN~\cite{garofalo2020pulp} library to support also Batch-Normalization and Quantization/Activation as defined in Eqs.~\ref{eq:6} and \ref{eq:7}, respectively, which together compose the Norm/Qnt phase.
The PULP-NN library assumes that all the activations and weights are stored in the L1 memory.
Readers may refer to \cite{garofalo2020pulp} for detailed information about this library.

\subsection{QNN Tensor Tiling}
%

In the context of QNN deployment, a tiling strategy consists of a regular software-managed fragmentation of the data tensors mentioned in Section~\ref{sec:qnns} to \textit{i}) fit within the available memory, and \textit{ii}) transparently move data between levels, using double buffering and DMA of the next tile in parallel with computation on the current tile.
In this work, we target a hardware architecture with three levels of memory hierarchy: a virtually unlimited-size off-chip L3; an on-chip L2 memory balancing size (e.g., 256~kB to a few MB) and bandwidth; and an on-chip L1 with virtually unlimited bandwidth to the compute units, but of limited size (typically $<$ 128~kB).

If we consider a convolutional layer in a DNN, in general, inputs, outputs, and weights should all be tiled to satisfy memory constraints at all levels $Li$ (see Table~\ref{tab:notation} for the notation adopted throughout the paper).
The main challenge of tiling is to maximize the size of all tiles while \textit{i}) fitting within the size constraints imposed by the size of layer $Li$, and \textit{ii}) guaranteeing that all relationships between the tensors are respected both on the tiles in $Li$ and on the full tensors in $L(i+1)$.

\section{DORY: Deployment Oriented to memoRY}
\label{sec:DORY}
DORY targets a compute node with three levels (L3, L2, and L1) in the memory hierarchy, as described in Section~\ref{sec:background}.
\rev{It supports L3-L2 and L2-L1 tiling of both weights and activations.
Storage of weights in L3 ($>$ 512 kB) is essential for the deployment of most non-trivial networks such as \cite{howard2017mobilenets,s2018mobilenetv2}.
On the other hand, activations' tiling is typically necessary only for networks working on high-resolution images with big spatial dimensions, which are rare in the edge computing domain.}
The operation of DORY is organized in three steps, performed offline before network deployment.
First, the \textit{ONNX decoder} receives as input a QNN graph using the Open Neural Network Exchange (ONNX format).
Then, the \textit{layer analyzer} optimizes and generates code to run the tiling loop, orchestrate layer-wise data movement and call a set of \textit{backend} APIs to execute each layer of the network, individually.
Finally, the \textit{network parser} merges information from the whole network to infer memory buffer sizes in each hierarchical level and orchestrate the end-to-end network execution.
It uses this information to generate an ANSI C file that embodies the whole DNN execution and can be compiled for the target platform.

\subsection{ONNX Decoder}
\label{sec:onnx}
The first operation performed by DORY is decoding the input ONNX graph representing an already quantized DNN, and reorganizing it in a set of layers.
In DORY, a \textit{layer} corresponds to a canonical sequence of operations performed by distinct ONNX graph nodes.
Each layer includes \textit{i}) a Linear/add/pooling operation, \textit{ii}) an optional Batch-Normalization operation, \textit{iii}) a Quantization/Activation operation.
Each DORY layer uses quantized inputs, outputs, and weight, while the representation of any temporary data is 32-bit signed integer.

\subsection{Layer Analyzer}
\label{sec:layerDORY}
\begin{figure*}
  \centering
\includegraphics[width=0.85\textwidth]{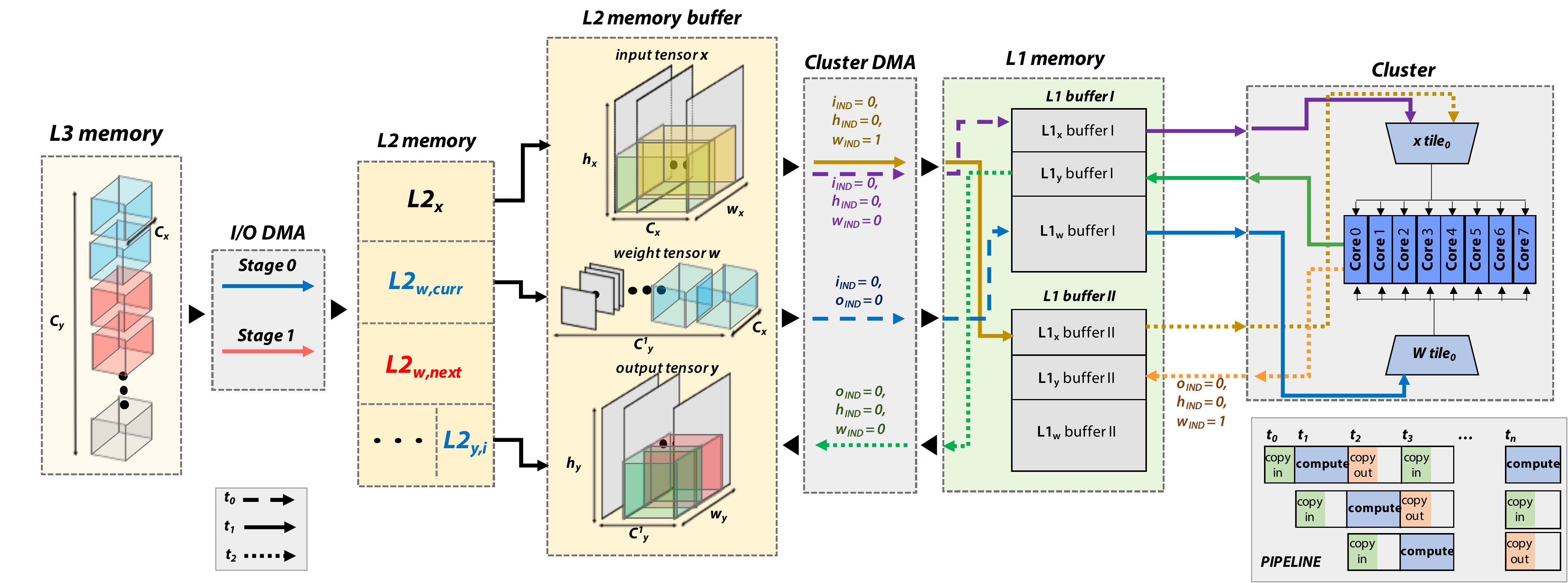}
  \caption{\rev{DORY L3-L2-L1 layer routine example. On the left, the I/O DMA copies weights tile in case only $C_y$ is L3-tiled. Two different buffers are used for $L2_w$.} Then, the Cluster DMA manages L2-L1 communication using double-buffering, while the cores compute a kernel on the current tile stored in one of the L1 buffers.}
\label{fig:L2_L1}
\vspace{-0.3cm}
\end{figure*}
\begin{listing}[tb]
\begin{minted}[mathescape,
              linenos,
              escapeinside=||,
              highlightlines={1,2,3,4},
              numbersep=5pt,
              gobble=2,
              frame=lines,
              fontsize=\footnotesize,
              framesep=2mm]{python}
  |\textcolor{blue}{\textbf{LTO}}|: for (o = 0; o < |$C_{y}^t$|; o++)  
    |\textcolor{blue}{\textbf{LTH}}|: for (h = 0; h < |$h_{y}^t$|; h  ++) 
      |\textcolor{blue}{\textbf{LTW}}|: for (w  = 0; w < |$w_{y}^t$|; w ++) 	
        |\textcolor{blue}{\textbf{LTI}}|: for (i = 0; i < |$C_{x}^t$|; i ++) 	
	  |\textcolor{darkspringgreen}{\textbf{dma\_wait}}($L1_{x,load}$)|;  |\textcolor{darkspringgreen}{\textbf{swap}}|(|$L1_{x,load}$|, |$L1_{x,exec}$|)
	  |\textcolor{darkspringgreen}{\textbf{dma\_async}}|(|$L1_{x,load}$| <- |$L2_x$|[i, w, h])
	  |\textcolor{darkspringgreen}{\textbf{dma\_wait}}|(|$L1_{w,load}$|);  |\textcolor{darkspringgreen}{\textbf{swap}}|(|$L1_{w,load}$|, |$L1_{w,exec}$|)
	  |\textcolor{darkspringgreen}{\textbf{dma\_async}}|(|$L1_{w,load}$| <- |$L2_w$|[i, o]) 	
	  if (o  + h + w  + i > 0)
		|\textcolor{darkspringgreen}{\textbf{DNN\_kernel}}| (|$L1_{x,exec}$|, |$L1_{w,exec}$|, |$L1_{y,exec}$|)
# from 3° iteration: fully operating pipeline
        if (o  + h  + w  + i > 1)
	  |\textcolor{darkspringgreen}{\textbf{dma\_wait}}|(|$L1_{y,load}$|)
	  |\textcolor{darkspringgreen}{\textbf{dma\_async}}|(|$L1_{y,load}$|  -> |$L2_y$|[ o, w, h])
	  |\textcolor{darkspringgreen}{\textbf{swap}}|(|$L1_{y,load}$| , |$L1_{y,exec}$|)
  
 \end{minted}
 \caption{DORY \rev{L2-L1} loop nest implementing the double buffering scheme as represented in right part of Figure \ref{fig:L2_L1}. At each most internal loop iteration, two asynchronous Cluster DMA calls are made to copy the weights and input activation of the next tile into L1 memory, the basic kernel is executed on the current tile, and one other cluster DMA transfer is executed to copy the output back on the L2 memory.} 
 \label{fig:pseudocode_layer}
\vspace{-0.4cm}
 \end{listing}

In the first optimization phase, DORY layers are considered separately from \fcrev{each other}, using only weight dimension information from the previous layer.
The layer analyzer includes three submodules: a platform-agnostic \textit{tiling solver}; \revTCOMP{a set of \textit{heuristics \& constraints} optimizing execution over a target-specific backend and limiting the tiling search space}; and a \textit{SW-cache generator}.
\subsubsection{DORY Tiling Solver} 
In the following discussion, we use the terminology defined in Section~\ref{sec:background} and denote a buffer residing in $Li$ memory as $Li_{t}$, where $\mathbf{t}$ is the name of the tensor. 
The Solver relies on a 2-step engine, which solves the L3-L2 tiling constrained problem \revTCOMP{first}, and the L2-L1 one \revTCOMP{afterwards}.
\revTCOMP{With L3-L2 tiling, we enable storing activations and weights in the L3 off-chip memory instead of the on-chip L2.
With respect to tools that do not support L3 tiling for activations, such as Tensorflow Lite Micro, this feature enables to support significantly larger layers.
}
The Solver verifies whether the layer memory occupation fits the L2 memory input constraint or needs to be stored in L3:
\begin{equation}
L2_{w,next}+L2_{w,curr}+L2_{x}+L2_y \overset{?}{<} L2\; .
\label{eq:6}
\end{equation}
\revTCOMP{
We search an L3 tiling solution using a five-stage cascaded procedure.
At each stage, we try to tile a different selection of buffers to fit the constraint of Eq.~\ref{eq:6}.
Whenever possible, the tiler tries to avoid L3-L2 tiling of output activations, which always requires a double number of transfers (from L2 to L3 when produced, and from L3 to L2 when consumed by another layer).
Instead, the tiler tries to keep output activations in L2 as much as possible.
If a stage satisfies Eq.~\ref{eq:6}, the L3-L2 Tiling Solver is stopped and the dimensions of tiles are saved. Otherwise, the next stage is tried.
%
%
\begin{enumerate}[leftmargin=1.3cm]
\item [\textit{stage 0.}] \texttt{L3-tile x, w, y = OFF, OFF, OFF}. If Eq.~\ref{eq:6} is directly satisfied, we proceed without L3-L2 tiling.
\item [\textit{stage 1.}] \texttt{L3-tile x = ON}.
This solution is selected when the output of the previous layer was tiled in L3, and therefore input tiling cannot be avoided.
Tiling is performed along the $h_x$ dimension of the input, to avoid 2D transfers at the L3-L2 interface.
The tiler splits the layer in a series of identical ones that work on a different stripe of the input image.
\item [\textit{stage 2.}] \texttt{L3-tile w = ON}. Weight tiling is enabled on the $C_{y}$ dimension, dividing the layer in a set of smaller layers that work on different channels of the output image with $C_{y}' < C_{y}$.
This solution can only be selected when the output of the previous layer is already in L2.
\item [\textit{stage 3.}] \texttt{L3-tile w , y = OFF, ON}. Weight tiling is disabled while output tiling is enabled: the approach is similar to input tiling, but requires doubling the DMA transfers for the tiled tensor across the full network execution.
\item [\textit{stage 4.}] \texttt{L3-tile w, y = ON,  ON}. The L3 tiling is enabled on both buffers, $y$, $weights$.
This solution is selected when no other solution can fit L2.
\end{enumerate}
}

After the L3 tiling step, the DORY solver processes the layer to find a suitable L2-L1 tiling scheme, which requires more effort due to the typically small sizes of L1 memories.
\revTCOMP{Compared to high-end computation engines, with much larger memories, a suboptimal sizing of the tensors for the L1 small MCUs memory can be even more detrimental in terms of performance, as exposed in Section~\ref{sec:tile_perf}.}
DORY abstracts this as a Constraint Programming (CP) problem, and exploits the CP solver from the open-source OR-Tools developed by Google AI~\footnote{https://developers.google.com/optimization/} to meet hardware and geometrical constraint (e.g., $C_y^t$ for output and weights must be the same), while maximizing an objective function.
The base objective function of the solver is to maximize L1 memory utilization:
\begin{equation}
\mathrm{max}(L1_{x} + L1_{y} + L1_{w})\; ,
\label{eq:7}
\end{equation}
manipulating the tile dimensions (e.g., $C_x^t$ and $C_y^t$).
The hardware constraint is related to the max L1 buffer dimensions:
$$
L1_{x} + L1_{y} + L1_{w} + L1_{backend} < \frac{L1}{2}\; .
$$
with $L1_{backend}$, the overhead of the backend kernel, such as the im2col memory occupation of PULP-NN backend~\cite{garofalo2020pulp} or any other support buffer (e.g., the intermediate full-precision accumulators for CHW based convolutions).
Topological and geometrical constraints are due to the relationships between each tensor's characteristic dimensions and other parameters of a layer; for example,
$$
h_{y}^t = \Big(h_{x}^t - (K_h - 1) + 2 \cdot p \Big)
$$
embodies the relationship between the height dimension in the output and the input tiles, with $p$ representing padding.

\subsubsection{\fcrev{Target-specific} Heuristics \& Constraints} 
\label{sec:heuristics}
To maximize performance, the objective function of Eq.~\ref{eq:7} can be augmented with a series of heuristics targeting a specific backend.
The heuristics are combined with the objective function of Eq.~\ref{eq:7} by means of a set of tweakable parameters:
\begin{equation}
    \mathrm{max}\Big( \alpha(L1_x+L1_y+L1_w) 
    + \sum_i\beta_i\mathcal{H}_i \Big) \; .
    \label{eq:objective}
\end{equation}
\revb{Here, we list four heuristics related to PULP-NN, the backend library exploited by DORY \fcrev{in our GAP-8 case study}}.
\begin{shortitem}
    \item {\texttt{HIDE\_IM2COL}}: the PULP-NN im2col buffer is reused for each output pixel; therefore, maximizing the number of output channels optimizes the reuse of input pixels, reducing the overhead to create the im2col:
    $$
    \mathcal{H}_{i2c} = C_y^t
    $$
    \item {\texttt{PAR\_BALANCE}}\footnote{The \texttt{PAR\_BALANCE} constraint is changed to $
    \mathcal{H}_{par} = (h_{y}^t \times w_y^t -1)\,\mathrm{mod}\,16$ for ``patological'' output activations with $h_y < 8$.}: PULP-NN divides workload among cores following primarily the $h$ dimension (i.e., a chunk of rows per core). Therefore, making this a multiple the number of cores (8) maximizes balance:
    $$
    \mathcal{H}_{par} = (h_{y}^t-1)\,\mathrm{mod}\,8
    $$
    \item {\texttt{MATMUL\_W}} and {\texttt{MATMUL\_CH}}: the innermost loop of PULP-NN is a 4x2 matrix multiplication on 4 output channels and 2 pixels in $w$ direction.
    Maximizing adherence of a tile to this scheme optimizes performance:
    $$
    \mathcal{H}_{mm\_w} = (w_{y}^t-1)\,\mathrm{mod}\,2\;;\;
    \mathcal{H}_{mm\_ch} = (C_{y}^t-1)\,\mathrm{mod}\,4
    $$
\end{shortitem}
Section~\ref{sec:tile_perf} discusses the effectiveness of the PULP-NN heuristics in delivering a good quality-of-results.
\rev{Additionally, Section~\ref{sec:tile_perf} describes the impact of applying these heuristics both to the main tiling problem and to the sizing of the layer borders tile.}

\revTCOMP{We impose an additional constraint to always perform a full computation along the channel direction:
$$
C_x^t = C_x\; 
$$
We choose not to tile the $C_x$ dimension to avoid the memory overhead of long-term storage (and therefore, transfer to $L2$ and $L3$) of 32-bit partially accumulated values produced by the backend.
For the same reason, we do not tile the spatial dimension of filters, i.e., $K_h$ and $K_w$.
While these constraints restrict the solution space, we observe that the purged solutions are sub-optimal.}

\subsubsection{DORY SW-cache Generator}
The SW-cache Generator is charged of automatically generating C code orchestrating the execution of a whole layer given the tiling solution found by the Tiling Solver.
It instantiates asynchronous data transfers and calls to the backend kernels, without any manual effort.
DORY uses a \textit{triple-buffering} approach for the communication between L3-L2 and L2-L1 memories: specifically, double-buffering is applied simultaneously between L3-L2 and L2-L1 (Figure \ref{fig:L2_L1}), and all data transfers are pipelined and asynchronous.
With this approach, we can almost completely hide the memory transfer overhead, as discussed in Section~\ref{sec:results}. 
While the code generator is necessarily not platform-agnostic, the approach we follow can be easily generalized to any computing node with a three-level memory hierarchy.

Listing~\ref{fig:pseudocode_layer} provides DORY's scheduling scheme of L2-L1 layer execution, through LTO, LTW, LTH, and LTI loops on output channels, height, width, input channels tiles, respectively.
Loop iteration limits are statically resolved by the \textit{DORY tiling Solver}.
Moreover, DORY autonomously controls the complete execution of the layer, by managing padding, stride, and overlap for every single tile (e.g., padding $>$ 0 for border tiles whereas padding = 0 for internal ones, when the input padding parameter is $>$ 0).
Using statically resolved parameters, we maximize the usage of immediates, reducing load/store operations inside the inner loops of the layer tiling.

The layer-wise loop nest detailed in Listing~\ref{fig:pseudocode_layer} and Fig.~\ref{fig:L2_L1} is executed in three concurrent pipeline stages: \textit{i}) a new computation starts and fill the output buffer that was not used in the previous cycle; \textit{ii}) the results of the last cycle are stored back in L2; \textit{iii}) a new set of inputs is loaded in L1. 
At each pipeline cycle, we swap the load and the execution buffer (\emph{swap} operation of Listing~\ref{fig:pseudocode_layer}) to enable double buffering. 

\subsection{\revb{DORY Hybrid Model}}
\label{sec:depthwise}

\begin{figure}
  \centering
\includegraphics[width=0.76\columnwidth]{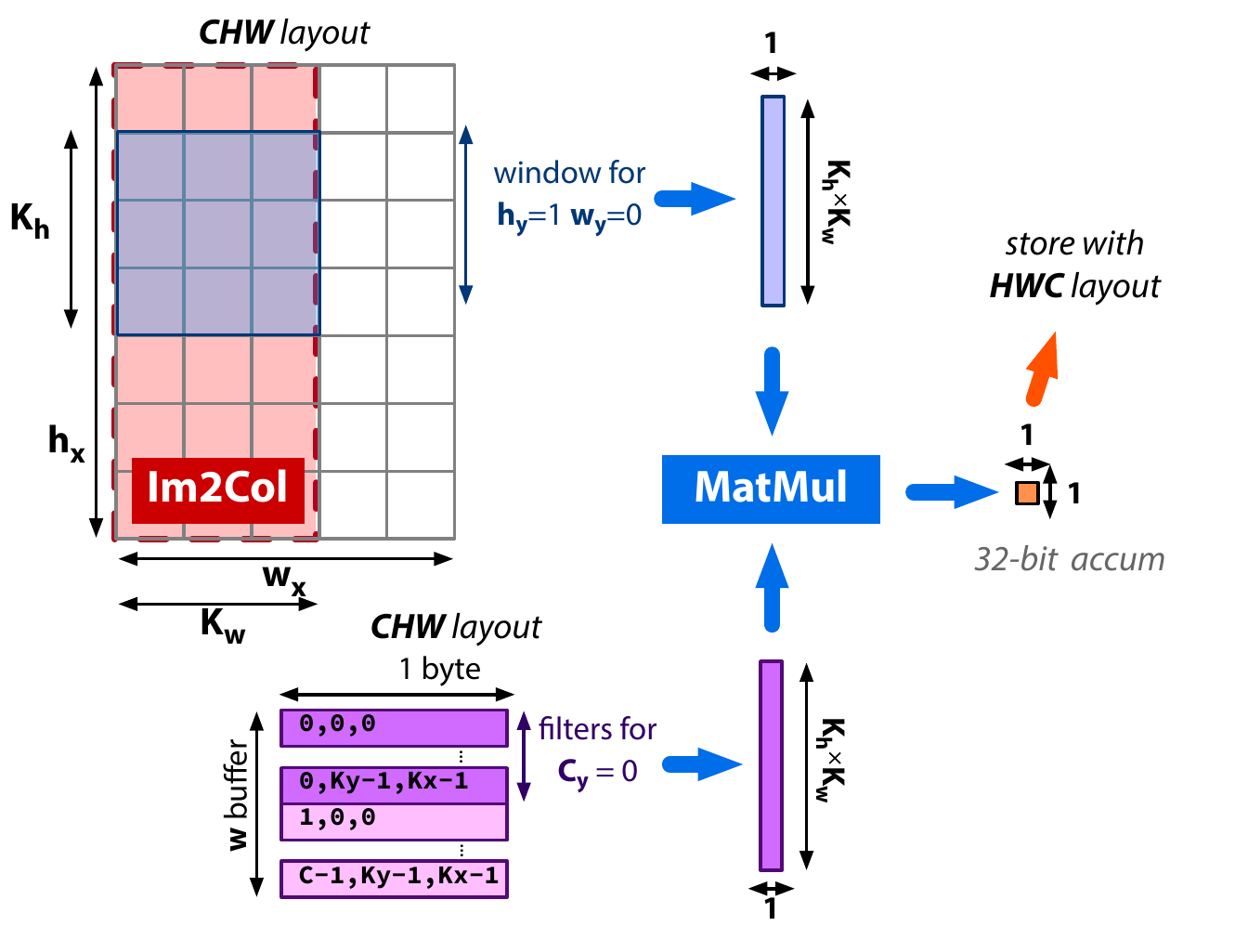}
\vspace{-0.5cm}
  \caption{Modified execution model for depthwise convolutions: the Im2Col buffer is built using a single channel out of CHW-layout activations; outputs are quantized and stored back using the PULP-NN model shown in Figure~\ref{fig:pulp_nn_overview} (see Table~\ref{tab:notation} for the buffer naming scheme).}
\label{fig:depthwise_overview}
\vspace{-0.5cm}
\end{figure}

\rev{
In the HWC data layout, used by CMSIS-NN~\cite{lai2018cmsis} and PULP-NN~\cite{garofalo2020pulp}, pixels referring to channels are contiguous, while spatially adjacent ones are stored with stride $>$ 1.
This layout enables constructing very optimized convolutional layers out of a single optimized matrix-multiplication kernel, by exploiting the reuse of activations over input channels \cite{lai2018cmsis, garofalo2020pulp} -- contrary to the CHW layout, which requires separately handcrafted and optimized kernels for each kernel size/stride configuration.
The main limit of this approach hits a specific category of convolutional layers, namely, depth-wise convolutions.
These do not accumulate over multiple channels; instead, they project each input channel into a single output channel disjointly from other channels. 
Therefore, they do not show any possibility to exploit channel data reuse. 
}

\rev{
On the one hand, depth-wise convolutions are unavoidable in modern networks for the edge, to decouple the channel-mixing and spatial filtering actions of the convolutional layer~\cite{s2018mobilenetv2}; on the other hand, they are typically only responsible for 10\% or less of the overall operations~\cite{howard2017mobilenets,s2018mobilenetv2}, meaning that directly optimizing for them may be suboptimal.
This scenario suggests a hybrid approach: using the HWC layout for general convolutional layers (and point-wise 1x1 layers), but switching to a hybrid CHW/HWC layout in depth-wise layers.
}


\rev{
Following this idea, we 
\revb{define new optimizations for existing layers and a new} depth-wise convolution that consumes and produces activations in HWC layout from L2/L3 memory, but reorders them in CHW layout on L1 to maximize the data reuse and, therefore, computational efficiency.
%
%
Specifically, multiple strided Cluster DMA transfers are used to marshal data from L2 converting it directly from the HWC to CHW layout.
An Im2Col buffer is constructed simply as a contiguous vertical stripe of width $K_w$; the innermost loop proceeds along the vertical stripe by computing a single output pixel per iteration.
The output pixels are then quantized and stored in an output buffer using the HWC layout, which can be directly transferred to L2.
Figure~\ref{fig:depthwise_overview} shows the execution model adopted for depthwise convolutions.
With this strategy, input data reuse -- the only kind available in depth-wise convolutions -- can be exploited along the vertical dimension, thanks to the fact that spatially adjacent pixels are contiguous in memory.
For parallel execution, multiple cores operate simultaneously on different channels; due to the channel independence, this choice minimizes memory contention, \fcrev{and optimizes performance while still keeping a degree of flexibility: the same kernel can be used to compute depth-wise layers of various filter shapes and strides}.
}

\subsection{Network Parser}
\label{sec:networkDORY}
%
%
\begin{listing}[tb]
\footnotesize
\begin{minted}[mathescape,
              linenos,
              escapeinside=||,
              highlightlines={3},
              numbersep=5pt,
              gobble=2,
              frame=lines,
              fontsize=\footnotesize,
              framesep=2mm]{python}
              
  |\textcolor{darkspringgreen}{\textbf{udma\_async}}|(|$L2_{w,load}$| <- |$L3_w$|[|$I_0$|])
  |\textcolor{darkspringgreen}{\textbf{udma\_wait}}|(|$L2_{w,load}$|);  	
  |\textcolor{blue}{\textbf{LTL}}|: for (i = 0; i < nlayers; i ++) 		                  
  # number of CNN layers 
    |\textcolor{darkspringgreen}{\textbf{udma\_wait}}|(|$L2_{w,load}$|); |\textcolor{darkspringgreen}{\textbf{swap}}|(|$L2_{w,load}$|, |$L2_{w,exec}$|)
    if  (layer{i+1} fit L2 && is Conv)
	|\textcolor{darkspringgreen}{\textbf{udma\_async}}|(|$L2_{w,load}$| <- |$L3_w$|[|$I_i$|])
    |\textcolor{darkspringgreen}{\textbf{Layer\{i\}}}| (|$L2_x$|,[|$L2_{x2}$|], [|$L3_w$|[|$I_i$|]], [|$L2_{w,exec}$|], |$L2_y$|)     
    # [] optional arguments 
	|\textcolor{darkspringgreen}{\textbf{swap}}|(|$L2_y$|, |$L2_x$|)	
    if  (layer{i} has residual) # bypass management
	|\textcolor{darkspringgreen}{\textbf{store}}|  (|$L2_y$|->|$L2_{x2}$|) 
    if  (layer{i} is Sum)
	|\textcolor{darkspringgreen}{\textbf{delete}}|  (|$L2_x2$|) 
	|\textcolor{darkspringgreen}{\textbf{Stack\_dealloc}}|(|$L2_y$|) # stack control
	|\textcolor{darkspringgreen}{\textbf{Stack\_alloc}}|(|$L2_x$|[|$I_{i+1}$|]) 

 \end{minted}
 
\vspace{-0.3cm}
  \caption{DORY network execution loop.}
\label{fig:pseudocode_nework}
\vspace{-0.4cm}
 \end{listing}
%
%
After layer-wise tiling has been completed by the Layer Analyzer, DORY uses the information extracted from all the layers to build a network graph, considering every single layer as a callable function.
Listing~\ref{fig:pseudocode_nework} showcases the execution loop of the DNN execution as created by our framework.
At each step, three main tasks are concatenated : \textit{i}) we transfer from L3 the weights of the following layer. \footnote{This phase is executed for layer \textit{i} only if layer \textit{i+1} is a convolution or a linear one and if it fits the dedicated space in the L2 memory. 
On the contrary, only the space for the $L2_w$ is allocated if the layer needs the L3-L2 tiling and no space at all is allocated if the layer \textit{i+1} is a pooling or an add.}
\textit{ii}) a new layer is executed pointing to the correct buffers inside the memory stack;
\textit{iii}) input and output buffer offsets are updated.

Similarly to single layers, the network-wise code is generated automatically without programmer intervention.
DORY produces a single function that \fcrev{can} be called inside a custom application \fcrev{by passing} two externally allocated memory buffers (for L1 and L2) and their maximum size as parameters.

\subsubsection{Buffer allocation stack \& Residual connections}
\label{sec:bidirectional_stack}
To allocate layer-wise input and output buffers in the L2 memory, we extend the two-stack strategy proposed by Palossi~et~al.~\cite{palossi201964mw}, employing a strategy based on a single bidirectional stack designed to avoid memory fragmentation and enable the execution of a sequence of differently sized layers.
%
Buffers are allocated/deallocated from the buffer allocation stack, which is constituted by two concurrent Last-In-First-Out stacks growing in opposite directions.
At the end of each layer's weight buffer allocation, we reverse the end of the stack for the next memory allocations.
By construction, the bidirectional stack is at worst as big as two concurrent stacks growing in the same direction.
For example, in a simple case without residual connections the dimension of our \textit{bidirectional stack} is 
$$
D_{stack} = \mathrm{max}_i (L2_{x,i}+L2_{w,i}+L2_{w,i+1}+L2_{x,i+1})\;,
$$ 
which is always less or equal than the size of two concurrent stacks $D_{stack,1}$, $D_{stack,2}$ due to the triangle inequality.

Before executing the $i$-th layer, the allocator manages the weight buffer $L2_{w,i}$ and output buffer $L2_{y,i}$; notice that $L2_{x,i}$ is already allocated as the $L2_{y,j}$ of a previously executed $j$-th layer (or the input of the network).
To manage residual connections, each $L2_{y,i}$ buffer has a \texttt{lifetime} counter associated.
To allocate a buffer in the stack for the $i$-th layer:
\begin{shortenum}
    \item one of the two corners of the stack is selected depending on a \texttt{begin\_end} flag that is switched at each new weight allocation;
    \item the allocator deallocates the last $L2_{w,i-2}$ buffer on the corner;
    \item the allocator checks if $L2_{y,i-2}$ has its \texttt{lifetime} counter set to 0; if so, it is deallocated;
    \item $L2_{y,i}$, $L2_{w,i}$ are allocated in order in the selected corner (with $L2_{w,i}$ nearest to the pointer);
    \item the \texttt{lifetime} counter of  $L2_{y,i}$ is set to the lifetime of the activation buffer, i.e., the number of layers to be executed before its deallocation.
    \item all \texttt{lifetime} counters are decreased by 1.
\end{shortenum}

The buffer allocation stack is naturally suited to execute a network with different branches (i.e., residual connections).
DORY always prioritizes the branch with the highest number of nodes.
The overall size of the stack is computed offline statically, taking into account all residual connections:
%
its dimension depends on the maximum sum of memory of two subsequent layers plus all the residuals from the previous layers.

\section{Results}
\label{sec:results}
In this section, we evaluate DORY in terms of quality-of-results (performance and energy efficiency) on both single layers and full networks, using \revb{GWT GAP-8 as a target platform for our exploration and our \fcrev{extended PULP-NN library} as a backend}.
We also compare our results with those obtained on a STM32-H743 MCU using STM X-CUBE-AI and on the same GAP-8 platform using the proprietary AutoTiler tool.
The results on single layers refer to a full 8-bit QNN layer as defined in Section~\ref{sec:qnns}, with Linear, Batch-Normalization, and Quantization/Activation sub-layers.
We set $\alpha$ to 0.5, $\beta_\texttt{HIDE\_IM2COL}$ to $10^2$, and other $\beta_i$ to $10^6$ in the objective function. 
\subsection{Single layer performance \& SoA comparison}
\label{sec:layer_perf}
\begin{figure}
  \centering
\includegraphics[width=0.92\columnwidth]{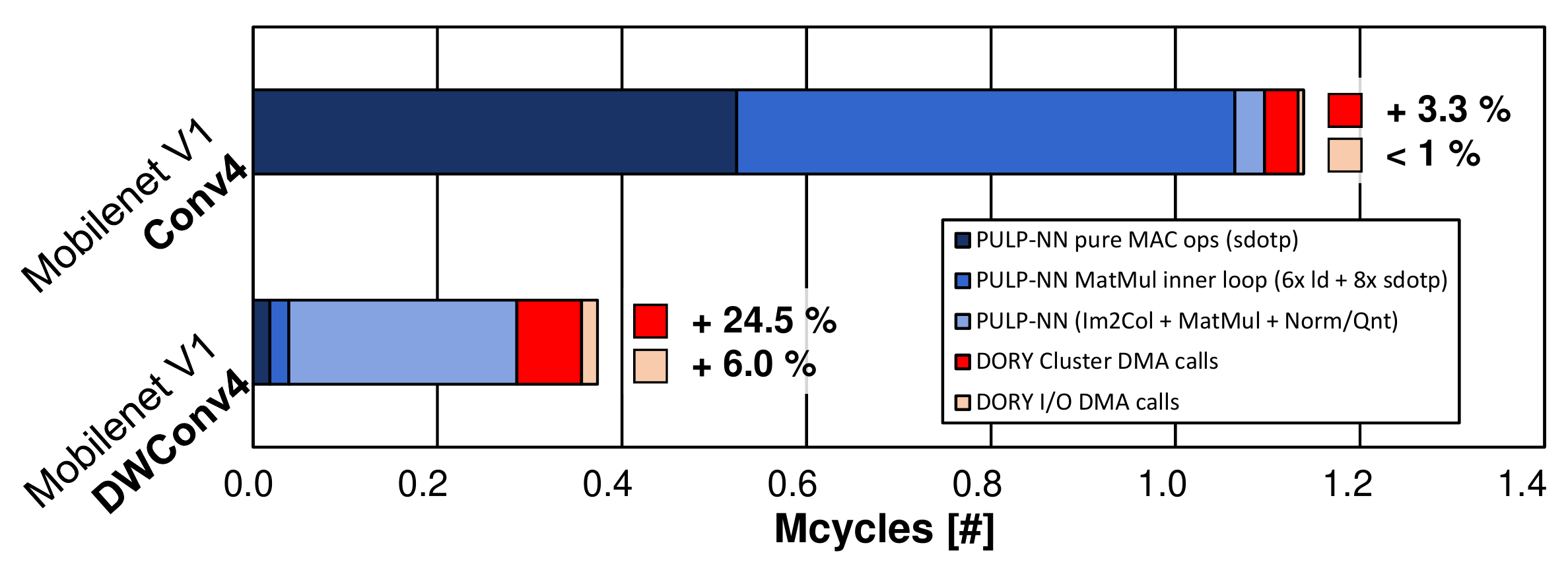}
\vspace{-0.5cm}
  \caption{\rev{Execution time analysis for point-wise and depth-wise layers.}}
\label{fig:degradation}
\vspace{-0.5cm}
\end{figure}
\begin{figure}[t]
  \centering
\includegraphics[width=0.92\columnwidth]{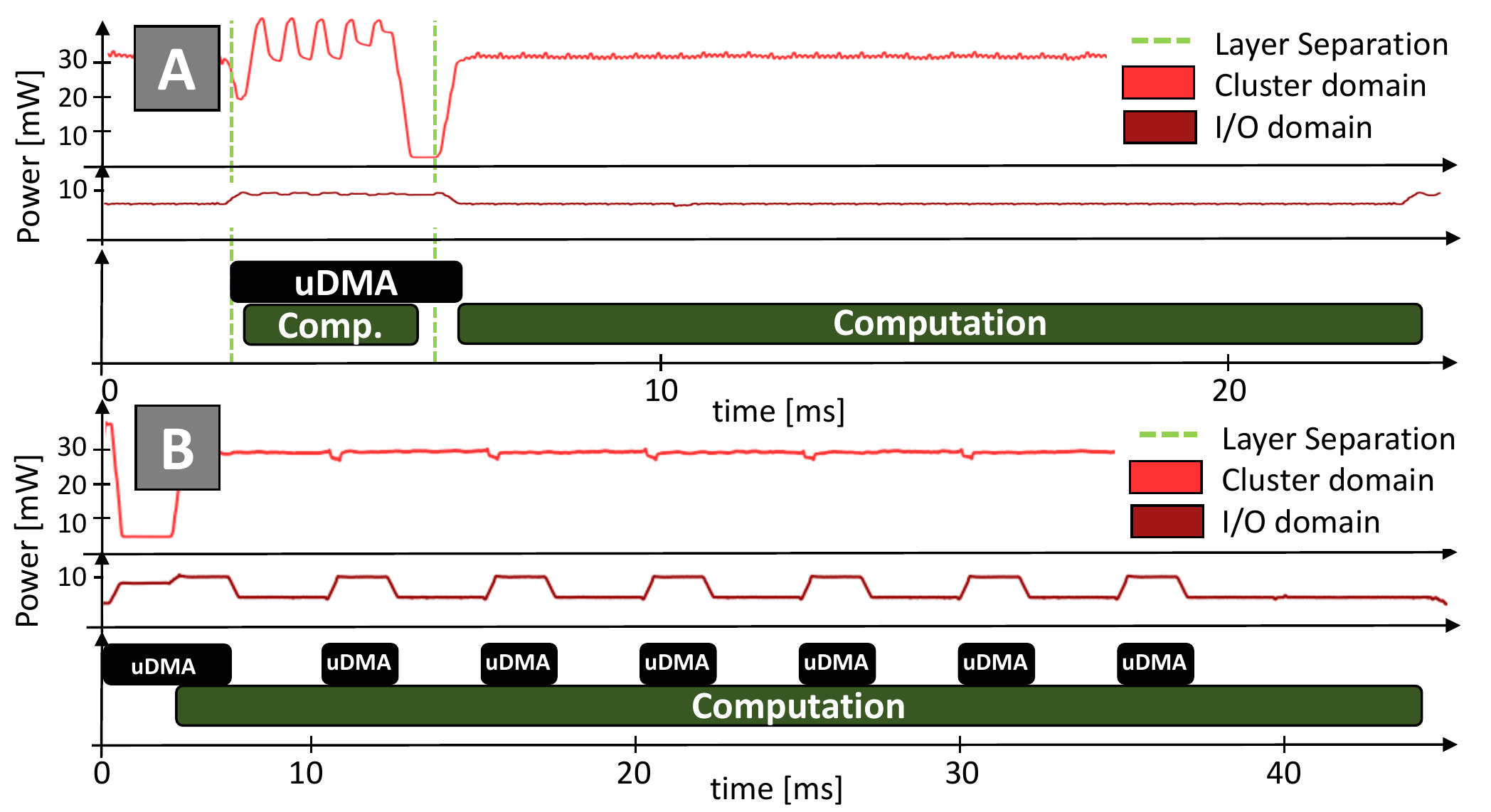}
\vspace{-0.3cm}
  \caption{In Part.A, the power traces of a point-wise Convolution following a depth-wise one. The I/O DMA causes the COREs to go in IDLE, waiting for the memory transfer end. In Part.B, an L3-tiled layer is executed and perfectly buffered to hide the memory hierarchy to the computing engine. fr = 100 MHz and $V_{DD}$ = 1V have been used on the GAP8 MCU. }
\label{fig:net_frags}
\vspace{-0.4cm}
\end{figure}
%
%
%
Fig.~\ref{fig:degradation} analyzes all the execution time for two layers of MobileNet-v1~\cite{howard2017mobilenets}, the first representative of point-wise convolutional layers, the second of depth-wise ones.
We observe several effects.
%
For the point-wise layer, roughly all the time is spent in the innermost loop of MatMul (most of which is pure MAC operations); the rest is due to building the Im2Col buffer, Norm/Qnt and MatMul loops that cover the SIMD leftover cases (e.g., $C_y^t$ not multiple of vector size 4).
In the case of depth-wise layers, this latter class of loops dominates the backend execution time.

\rev{
For what concerns the overhead introduced by DORY-generated tiling, we observe that the Cluster DMA does not impair the point-wise convolutional layers, since they are compute-bound and efficiently pipelined:
further, the processing overhead of calling the Cluster DMA many times is parallelized over the 8 cores in the cluster, reducing the cost of realizing complicated tiling schemes.
On the other hand, depth-wise layers are small, and both the Cluster DMA and I/O DMA overheads are exacerbated. Therefore, the load of the internal tiles and the asynchronous I/O DMA load of the following layer's weights are often impacting performance.
Fig.~\ref{fig:net_frags} corroborates this conclusion, showing power valleys in the cluster computation while waiting for new tile transfers (smaller ones) and for the weights of next layers to be transferred from the external memory.
Instead, Part.B of Fig.~\ref{fig:net_frags} shows the execution of a point-wise L3-tiled layer: in this case, the computation perfectly overlaps with the memory transfers, completely hiding the memory transfer overhead.}

\begin{table}[t]
\centering
{\color{black}
\begin{threeparttable}
\caption{Average performance and efficiency on 8-bits MobileNet-V1 layers obtained with DORY and other SoA MCU-deployment frameworks.}
\label{tab:Soa-layer}
{
\centering
\footnotesize
\renewcommand{\arraystretch}{1.2}
\setlength\tabcolsep{5pt}
\begin{tabulary}{\columnwidth}{CC|C|C|C}
                                                                   &         & \multicolumn{2}{c|}{\textbf{Performance (speed-up)}}        & \textbf{Efficiency}  \\                                    &         &  MAC/cycle             &   GMAC/s               &   GMAC/s/W             \\ \hline\hline
\multicolumn{1}{c|}{\multirow{2}{*}{\begin{tabular}[c]{@{}c@{}}\textbf{TFLite}\\ \textbf{Micro}\end{tabular}\tnote{a}}} & DwConv & 0.064 (0.2$\times$) & 0.03 (0.2$\times$)  & 0.13 (0.2$\times$)     \\ \cline{2-5} 
\multicolumn{1}{c|}{}                                                                              & PwConv    & 0.056 (0.1$\times$) & 0.027 (0.1$\times$) & 0.11 (0.1$\times$)      \\ \hline\hline
\multicolumn{1}{c|}{\multirow{2}{*}{\begin{tabular}[c]{@{}c@{}}\textbf{STM}\tnote{a}\\\textbf{CUBE-AI}\end{tabular}}}        & DwConv & 0.39 (1$\times$)     & 0.19 (1$\times$)     & 0.8 (1$\times$) \\ \cline{2-5} 
\multicolumn{1}{c|}{}                                                                              & PwConv    & 0.71 (1$\times$)     & 0.34 (1$\times$)     & 1.46 (1$\times$)  \\ \hline\hline

\multicolumn{1}{c|}{\multirow{2}{*}{\begin{tabular}[c]{@{}c@{}}\textbf{GWT}\tnote{b}\\ \textbf{AutoTiler}\end{tabular}}}    & DwConv & 2.16 (5.5$\times$)   & 0.22 (1.2$\times$)   & 4.24 (5.3$\times$)   \\ \cline{2-5} 
\multicolumn{1}{c|}{}                                                                              & PwConv    & 7.87 (11.1$\times$)  & 0.79 (2.3$\times$)   & 15.4 (10.6$\times$)  \\ \hline\hline

\multicolumn{1}{c|}{\multirow{2}{*}{\begin{tabular}[c]{@{}c@{}}\textbf{GWT}\tnote{c}\\ \textbf{AutoTiler}\end{tabular}}}    & DwConv & 2.16 (5.5$\times$)   & 0.56 (3.0$\times$)   & 2.16 (2.7$\times$)  \\ \cline{2-5} 
\multicolumn{1}{c|}{}                                                                              & PwConv    & 7.87 (11.1$\times$)  & 2.05 (6.0$\times$)   & 7.87 (5.4$\times$) \\ \hline\hline

\multicolumn{1}{c|}{\multirow{2}{*}{\textbf{DORY}\tnote{b}}}                                                         & DwConv & 1.14 (2.9$\times$)   & 0.11 (0.6$\times$)   & 2.24 (2.8$\times$)  \\ \cline{2-5} 
\multicolumn{1}{c|}{}                                                                              & PwConv    & 12.86 (18.1$\times$) & 1.29 (3.8$\times$)   & 25.2 (17.3$\times$)\\\hline\hline

\multicolumn{1}{c|}{\multirow{2}{*}{\textbf{DORY}\tnote{c}}}                                                         & DwConv & 1.14 (2.9$\times$)   & 0.30 (1.6$\times$)   & 1.14 (1.4$\times$)  \\ \cline{2-5} 
\multicolumn{1}{c|}{}                                                                              & PwConv    & 12.86 (18.1$\times$) & 3.34 (9.8$\times$)   & 12.86 (8.8$\times$)    \\\hline\hline

\end{tabulary}
}
\begin{tablenotes}
\item [a] Collected on the STM32H743 @ 480MHz.
\item [b] Collected on the GWT GAP8 @ (100MHz, 1V).
\item [c] Collected on the GWT GAP8 @ (260MHz, 1.15V).
\end{tablenotes}
\end{threeparttable}}
\vspace{-0.5cm}
\end{table}

%
\begin{figure*}
  \centering
\includegraphics[width=0.88\textwidth]{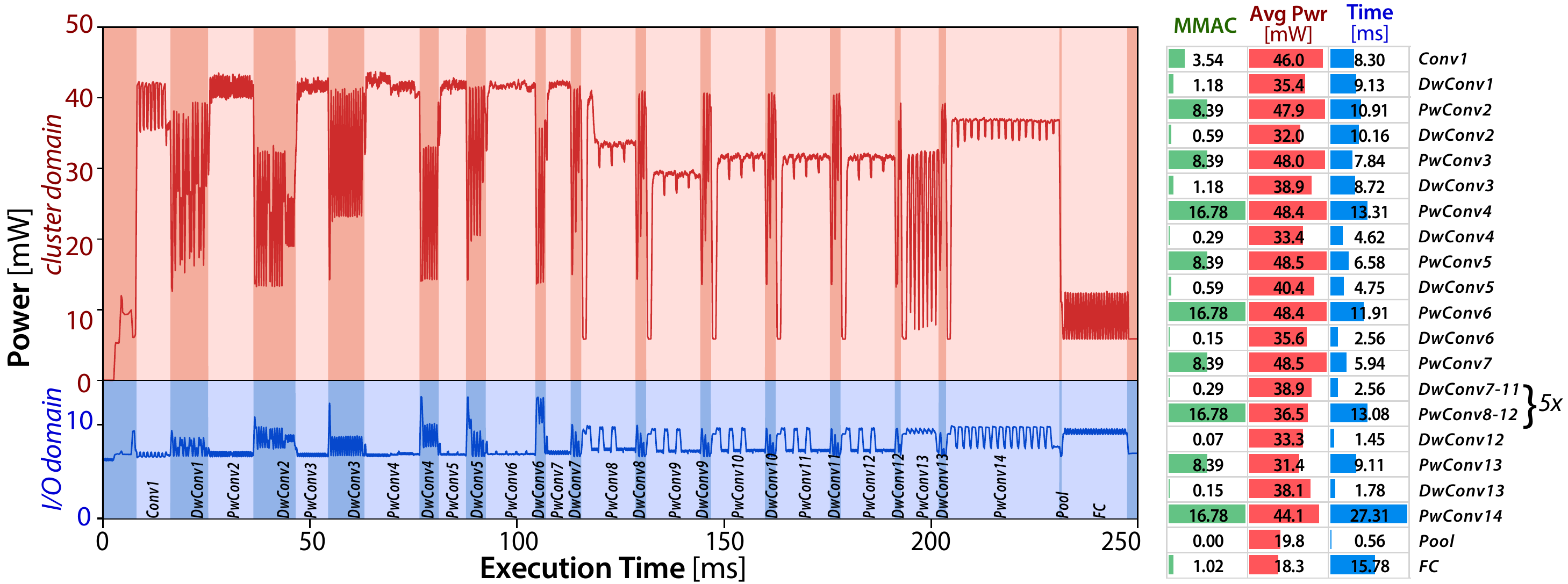}
\vspace{-0.3cm}
  \caption{In the left part, the 1.0-MobileNet-128 power profile when running on GAP-8 @ $f_\mathrm{cluster}=f_\mathrm{io}=\SI{100}{\mega\hertz}$ and $V_{DD}=\SI{1}{\volt}$. On the right, number of MAC operations, average power, and time for each layer of the network. Power was sampled at 64 KHz and then filtered with a moving average of \SI{300}{\micro\second}.}
\label{fig:net_energy}
\vspace{-0.5cm}
\end{figure*}

In Table~\ref{tab:Soa-layer}, we compare our approach with three state-of-the-art frameworks for DNN deployment on MCU: TFLite Micro, STM X-CUBE-AI, and GWT AutoTiler.
We focus on convolutional and depth-wise convolutional layers, which constitute the vast majority of computation in modern DNN models.
Metrics are computed as the average between the layers in the MobileNet-v1 network.
Results obtained on TFLite Micro and STM X-CUBE-AI refer to an STM32H743 microcontroller, based on an ARM Cortex-M7; those for GWT AutoTiler and DORY to a GWT GAP-8, described in Section~\ref{subsec:pulp}.
All results refer to 8-bit quantized networks, even if STM32 also supports 32-bit floating point; accuracy is equivalent to that of a non-quantized network.

TFLite Micro has the main advantage of being available on many different ARM and RISC-V MCUs; on the other hand, its performance is severely limited by the fact that it uses very general APIs without deep optimizations.
X-CUBE-AI outperforms it by 6.1$\times$ to 12.7$\times$ on the same platform, thanks to its much more efficient backend.
\revTCOMP{Nonetheless, layers generated by DORY for the GAP-8 platform outperform both TFLite Micro and X-CUBE-AI by a margin of 2.9$\times$ to 229.6$\times$ in terms of MAC/cycle.
This significant advantage is due to the architectural benefits of GAP-8 (multi-core acceleration, DSP-enhanced instructions) that DORY can exploit fully through PULP-NN, as showcased in the previous section.
In Section~\ref{sec:DORY_single_core}, we decouple DORY performance enhancement and architectural benefits to underline the benefits of our framework, deploying layers with DORY both on the STM32H7 and on GAP8 forced to run with a single-core. 
}

When compared to GWT AutoTiler, which targets the same platform, DORY is \rev{1.6$\times$} faster in point-wise convolutions, while it pays a performance toll in depth-wise convolutions, where it is \rev{1.9$\times$ slower}.
\rev{
These differences amount mainly to the different strategies followed by the tools in their respective backends and will be deeply discussed in Section~\ref{subsec:layout}.
}
As explained in Section~\ref{subsec:pulp}, the number of output channels strongly influences performance because re-using input data for more output channels offsets the cost of the Im2Col operation.
For depth-wise convolutions, each input channel is linked to a single output channels: as a consequence, this source of data re-use is not available.
%
%

\subsection{End-to-end network performance}
\label{sec:DNN_performance}
\begin{table*}[t]
\caption{End-to-end execution of image recognition MobileNet-v1 and MobileNet-v2 on GAP8 and STM32H7 MCUs.
}
\vspace{-0.3cm}
\label{tab:comparison}
\centering
\renewcommand{\arraystretch}{1.2}
\revTCOMP{
\resizebox{\textwidth}{!}{
\begin{tabulary}{\textwidth}{LCCCCCCCCCCCC}
\multicolumn{1}{c}{\multirow{2}{*}{\textbf{Configuration}}} & \multicolumn{1}{c}{\multirow{2}{*}{\textbf{Params}}}  & \multicolumn{1}{c}{\textbf{Work}}    & \multicolumn{1}{c}{\multirow{2}{*}{\textbf{Cycles}}}  & \multicolumn{1}{c}{\textbf{Perf}} & \multicolumn{1}{c}{\textbf{Eff}} & \multicolumn{1}{c}{\textbf{Perf}} & \multicolumn{1}{c}{\textbf{Lat}} & \multicolumn{1}{c}{\textbf{Energy}} & \multicolumn{1}{c}{\textbf{Eff}} & \multicolumn{1}{c}{\textbf{Perf}} & \multicolumn{1}{c}{\textbf{Lat}} & \multicolumn{1}{c}{\textbf{Energy}} \\
\multicolumn{1}{c}{}              & & \multicolumn{1}{c}{MAC}    & & \multicolumn{1}{c}{MAC/cyc} & \multicolumn{1}{c}{GMAC/s/W} & \multicolumn{1}{c}{GMAC/s} & \multicolumn{1}{c}{lat. {[}ms{]}} & \multicolumn{1}{c}{E {[}mJ{]}} & \multicolumn{1}{c}{GMAC/s/W} & \multicolumn{1}{c}{GMAC/s} & \multicolumn{1}{c}{lat. {[}ms{]}} & E {[}mJ{]} \\ \hline \hline
\multicolumn{13}{c}{\textit{DORY} @ GAP8}                    \\ \hline \hline
\multicolumn{5}{l|}{}       & \multicolumn{4}{|c|}{Low energy 1V @ 100 MHz}                                                                                       & \multicolumn{4}{|c}{Low latency 1.15V @ 260 MHz}                                                              \\ \hline
\multicolumn{1}{l}{1.0-M.V1-128}  & \multicolumn{1}{l}{4.2 M}  & \multicolumn{1}{l}{186.4 M} & \multicolumn{1}{l}{23.3 M}  & \multicolumn{1}{l|}{8.00}     & \multicolumn{1}{|l}{15.68}    & \multicolumn{1}{l}{0.80}   & \multicolumn{1}{l}{233.11}        & \multicolumn{1}{l|}{11.89}      & \multicolumn{1}{|l}{7.93}     & \multicolumn{1}{l}{2.08}   & \multicolumn{1}{l}{89.66}        & 23.51      \\ \hline
\multicolumn{1}{l}{0.5-M.V1-192}  & \multicolumn{1}{l}{1.3 M}  & \multicolumn{1}{l}{110.0 M} & \multicolumn{1}{l}{16.0 M}  & \multicolumn{1}{l|}{6.86}     & \multicolumn{1}{|l}{13.46}    & \multicolumn{1}{l}{0.69}   & \multicolumn{1}{l}{160.2}        & \multicolumn{1}{l|}{8.17}      & \multicolumn{1}{|l}{6.82}     & \multicolumn{1}{l}{1.78}   & \multicolumn{1}{l}{61.62}        & 16.16      \\ \hline
\multicolumn{1}{l}{0.25-M.V1-128}  & \multicolumn{1}{l}{0.5 M}  & \multicolumn{1}{l}{13.5 M} & \multicolumn{1}{l}{2.8 M}  & \multicolumn{1}{l|}{4.74}     & \multicolumn{1}{|l}{9.30}    & \multicolumn{1}{l}{0.47}   & \multicolumn{1}{l}{28.50}        & \multicolumn{1}{l|}{1.45}      & \multicolumn{1}{|l}{4.69}     & \multicolumn{1}{l}{1.23}   & \multicolumn{1}{l}{10.95}        & 2.87      \\ \hline
\multicolumn{1}{l}{1.0-M.V2-128}  & \multicolumn{1}{l}{3.47 M} & \multicolumn{1}{l}{100.1 M}  & \multicolumn{1}{l}{19.0 M}  & \multicolumn{1}{l|}{5.27}     & \multicolumn{1}{|l}{10.33}     & \multicolumn{1}{l}{0.53}   & \multicolumn{1}{l}{190.03}        & \multicolumn{1}{l|}{9.69}      & \multicolumn{1}{|l}{5.22}     & \multicolumn{1}{l}{1.37}   & \multicolumn{1}{l}{73.09}         & 19.16      \\ \hline 
\hline
\multicolumn{13}{c}{\textit{GWT AutoTiler} @ GAP8}   \\ \hline \hline
\multicolumn{5}{l|}{}        & \multicolumn{4}{|c|}{Low energy 1V @ 100 MHz}                                                                                       & \multicolumn{4}{|c}{Low latency 1.15V @ 260 MHz}                                                              \\ \hline
\multicolumn{1}{l}{1.0-M.V1-128}  & \multicolumn{1}{l}{4.2 M}  & \multicolumn{1}{l}{186.4 M} & \multicolumn{1}{l}{28.1 M}  & \multicolumn{1}{l|}{6.64}     & \multicolumn{1}{|l}{13.02}     & \multicolumn{1}{l}{0.66}   & \multicolumn{1}{l}{280.80}        & \multicolumn{1}{l|}{14.32}      & \multicolumn{1}{|l}{6.58}     & \multicolumn{1}{l}{1.73}   & \multicolumn{1}{l}{108.00}        & 28.32      \\ \hline 
\multicolumn{1}{l}{1.0-M.V2-128}  & \multicolumn{1}{l}{3.47 M} & \multicolumn{1}{l}{100.1 M}  & \multicolumn{1}{l}{19.7 M}  & \multicolumn{1}{l|}{5.07}     & \multicolumn{1}{|l}{9.95}     & \multicolumn{1}{l}{0.51}   & \multicolumn{1}{l}{197.38}        & \multicolumn{1}{l|}{10.07}      & \multicolumn{1}{|l}{5.03}     & \multicolumn{1}{l}{1.32}   & \multicolumn{1}{l}{75.92}         & 19.91     \\
\hline\hline
\multicolumn{13}{c}{\textit{X-CUBE-AI} @ STM32H7, solutions fitting 2MB ROM + 512 kB R/W RAM @ 480 MHz \cite{capotondi2020cmix}}   \\ \hline \hline
\multicolumn{1}{l}{0.25-M.V1-128} & \multicolumn{1}{l}{0.5 M}  & \multicolumn{1}{l}{13.5 M}  & \multicolumn{1}{l}{26.0 M}  & \multicolumn{1}{l|}{0.52}     & \multicolumn{1}{|l}{1.07}     & \multicolumn{1}{l}{0.25}   & \multicolumn{1}{l}{51.14}         & \multicolumn{1}{l|}{12.67}      & \multicolumn{1}{|l}{n.a.}     & \multicolumn{1}{l}{n.a.}   & \multicolumn{1}{l}{n.a.}          & n.a.       \\ \hline
\multicolumn{1}{l}{0.5-M.V1-192}  & \multicolumn{1}{l}{1.37 M} & \multicolumn{1}{l}{109.5 M} & \multicolumn{1}{l}{212.3 M} & \multicolumn{1}{l|}{0.52}     & \multicolumn{1}{|l}{1.06}     & \multicolumn{1}{l}{0.25}   & \multicolumn{1}{l}{442.27}        & \multicolumn{1}{l|}{103.49}     & \multicolumn{1}{|l}{n.a.}     & \multicolumn{1}{l}{n.a.}   & \multicolumn{1}{l}{n.a.}          & n.a.       \\ \hline
\end{tabulary}
}
\vspace{-0.5cm}
}
\end{table*}
In this Section, we focus on the performance of DORY in deployment full-networks that are already used as benchmarks for many edge-oriented works~\cite{capotondi2020cmix}.
%
%
%
\revTCOMP{
All the networks were run on GWT GAP-8, verifying all intermediate results as well as the final result of end-to-end runs against a PyTorch-based bit-accurate golden model for QNNs~\cite{conti2020nemo}, to confirm the correct functionality of the DORY framework and the PULP-NN backend.
}

\revTCOMP{
\subsubsection{End-to-end MobileNet-v1 and -v2 \& SoA comparison}
}
%

%
%
%
%
%
%
\revTCOMP{
Table~\ref{tab:comparison} showcases a full comparison in terms of energy efficiency (GMAC/s/W), throughput (GMAC/s), latency, and energy per frame.
%
%
Different variations of the MobileNet-v1 have been compared, with the same topology but a different number of channels or input dimensions. 
For state-of-the-art, we show the biggest networks that fit the on-chip/off-chip memory of the STM32H7 and GAP8, respectively (compatible with the ones deployed with DORY).
We will release similar benchmarks on all the Mobilenets on our public repository.
%
%
As can be noticed from the Table, DORY on MobileNet-v1 achieves up to 13.19$\times$ higher throughput in MAC/cycles than the execution on an STM32H7 (on 0.5-M.V1-192), using the best framework (X-CUBE-AI) currently available.
On different operating points, we have up to 7.1$\times$ throughput (1.78 vs. 0.25 GMAC/s) and 12.6$\times$ better energy efficiency, given the different frequencies and power consumption of the two platforms.
%
%
}

\revTCOMP{
Compared with GWT-proprietary and partially closed-source AutoTiler run on the same GAP-8 platform, our results show that DORY performs on average 20.5\% better.
}
Moreover, we note that as a default execution model, GWT AutoTiler folds the BatchNormalizations inside the convolution transformation, by saving operations, but potentially leading to more severe accuracy loss.
Contrarily to the Autotiler, DORY \fcrev{by default keeps} the BN \fcrev{explicit}, \revb{causing 3 extra LOADS and 1 additional MAC for each output pixel.} 
When \fcrev{using a 1:1 identical network to the one used by GWT AutoTiler (including folding)}, the performance gain is further increased to 26.6\%.
As previously discussed, the advantage lies in \textit{1)} the more efficient backend (PULP-NN) and \textit{2)} the heuristics, which guarantee that the tiling solution is optimized for the PULP-NN execution model.

\subsubsection{In-depth analysis of MobileNet-v1 execution}
Fig.~\ref{fig:net_energy} depicts the power profile of the end-to-end execution of a MobileNet-v1 (1.0 width multiplier, $128\times 128$ resolution) on GAP-8, with both the cluster and the fabric controller running at \SI{100}{\mega\hertz}.
The power consumption of the cluster domain (including 8 RI5CY cores, the L1 and the Cluster DMA) and of the I/O domain (including 1 RI5CY core, the L2, and the I/O DMA) is shown separately in two separate subplots.
In the cluster domain, power is dominated by the cores when the computation is in the active phase. 
Small valleys within a layer are given by (short) waits for the end of a memory transfer where the cores are all idle, or by Cluster DMA calls where a single core is active.
%
In the I/O domain, we can notice the I/O DMA consumption spikes: at the beginning of each layer, the weights of the following one are transferred from L3 to L2.

\section{Ablation Study}
\label{sec:abl_study}
This section presents a detailed ablation study of each of our contributions against state-of-the-art baselines.
\revTCOMP{We separately analyze the impact of: \textit{i)} the proposed heuristics; \textit{ii)} the hybrid optimization for depthwise layers;  \textit{iii)} voltage and frequency scaling on GAP-8; \textit{iv)} the size of L1 and L2 memories;  \textit{v)} the specific GAP-8 architecture compared to standard MCUs.}
%
%
\subsection{Single tile performance}
\label{sec:tile_perf}
\begin{figure}
  \centering
\includegraphics[width=0.9\columnwidth]{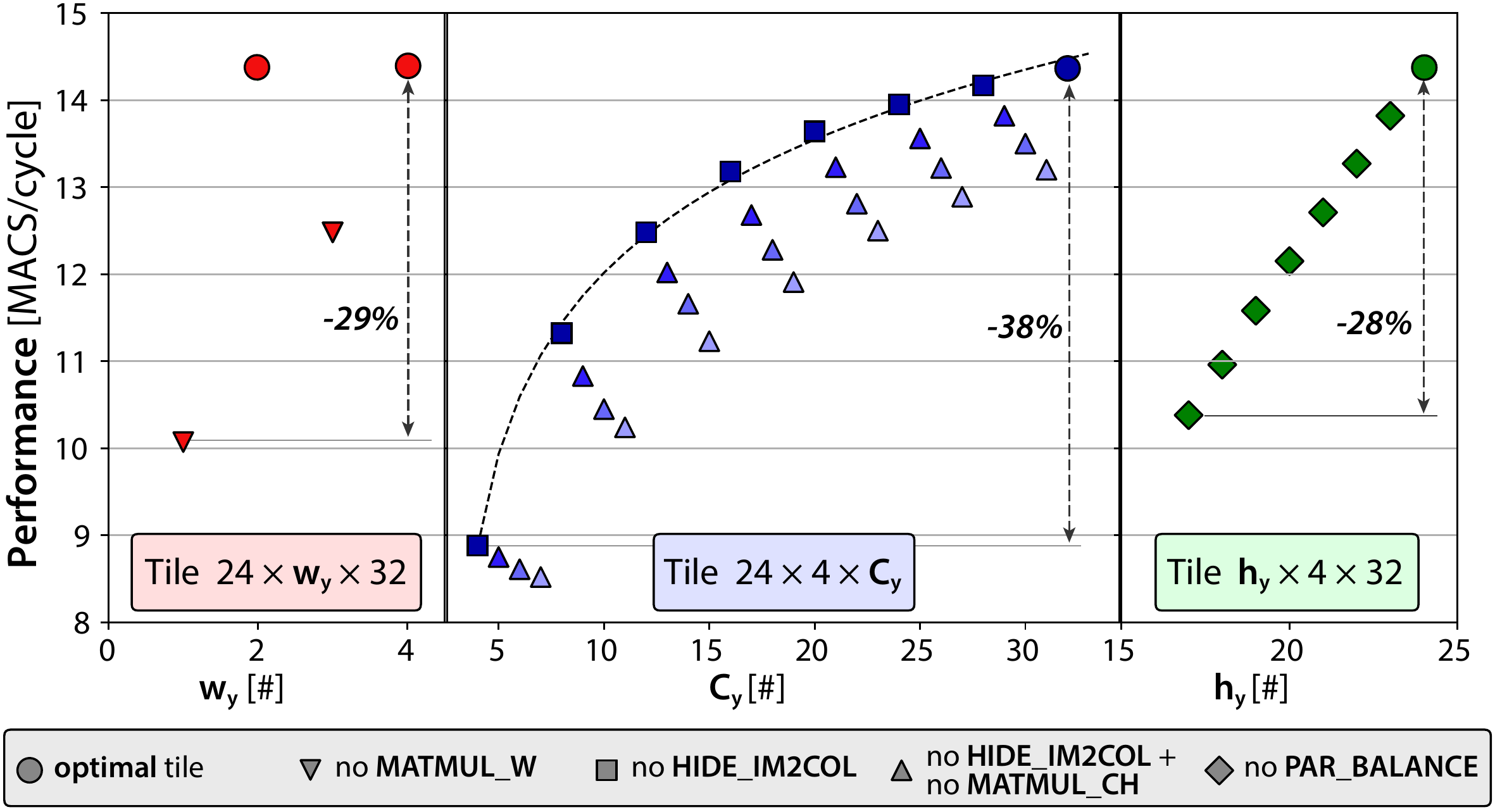}
\vspace{-0.2cm}
  \caption{Example of the effect of heuristic optimizations on convolutional layer performance. In this case, the ``optimal'' tile has output tensor 24$\times$4$\times$32 (HWC) and weight tensor 32$\times$3$\times$3$\times$32 (CoHWCi). Different optimizations are showed by varying $w_{y}$, $h_{y}$, and $C_{y}$ and violating the heuristics of Section~\ref{sec:heuristics}. }
\label{fig:opt}
\vspace{-0.6cm}
\end{figure}
We analyze the effects that the heuristics proposed in Section~\ref{sec:heuristics} have on the quality-of-results of the tiling solution. 
Moreover, we show the effect of applying these techniques to the border tile, increasing the performance in different configurations.
In particular, the size of the tile influences the execution efficiency of the backend layer. As such, a sub-optimal tiling choice can significantly reduce performance in the execution of a single inner tile.
Figure~\ref{fig:opt} exemplifies this phenomenon starting from an ``optimal'' tile of output tensor $24\times4\times32$ (HWC) with a $32\times3\times3\times32$ filter (channel out - height - width - channel in, or CoHWCi).
Violating \texttt{MATMUL\_W/CH} leads to a maximum performance loss of 29\%, violation of \texttt{HIDE\_IM2COL} to a 38\% loss, and violation of \texttt{PAR\_BALANCE} to a 28\% loss in this example layer.
Note that the performance loss is cumulative since each heuristic is written to improve the performance of a different section of the PULP-NN kernel.

To further underline this effect, if we set all the $\beta_i$ coefficients into the objective function of Eq.~\ref{eq:objective} to 0 and only focus on the maximization of the tile sizes, DORY chooses a tiling scheme that achieves only 2.78 MAC/cycles, 80.6\% lower than the 14.37 MAC/cycles achieved with the $\beta_i$ values previously reported.
In fact, in contrast with a superficial intuition, border tiles can constitute up to 50\% of the workload: for a layer of dimension 32$\times$64$\times$64$\times$32 (CoHWCi), the DORY tiler generates a main 32$\times$56$\times$2$\times$32 tile and a border 32$\times$8$\times$2$\times$32 tile with a 28~kB L1 memory constraint; both tiles are executed 32 times.
%
%
%
%

%
\revTCOMP{\subsection{Hybrid optimization for Depthwise layers}}
\label{subsec:layout}
\begin{figure}
  \centering
\includegraphics[width=0.85\columnwidth]{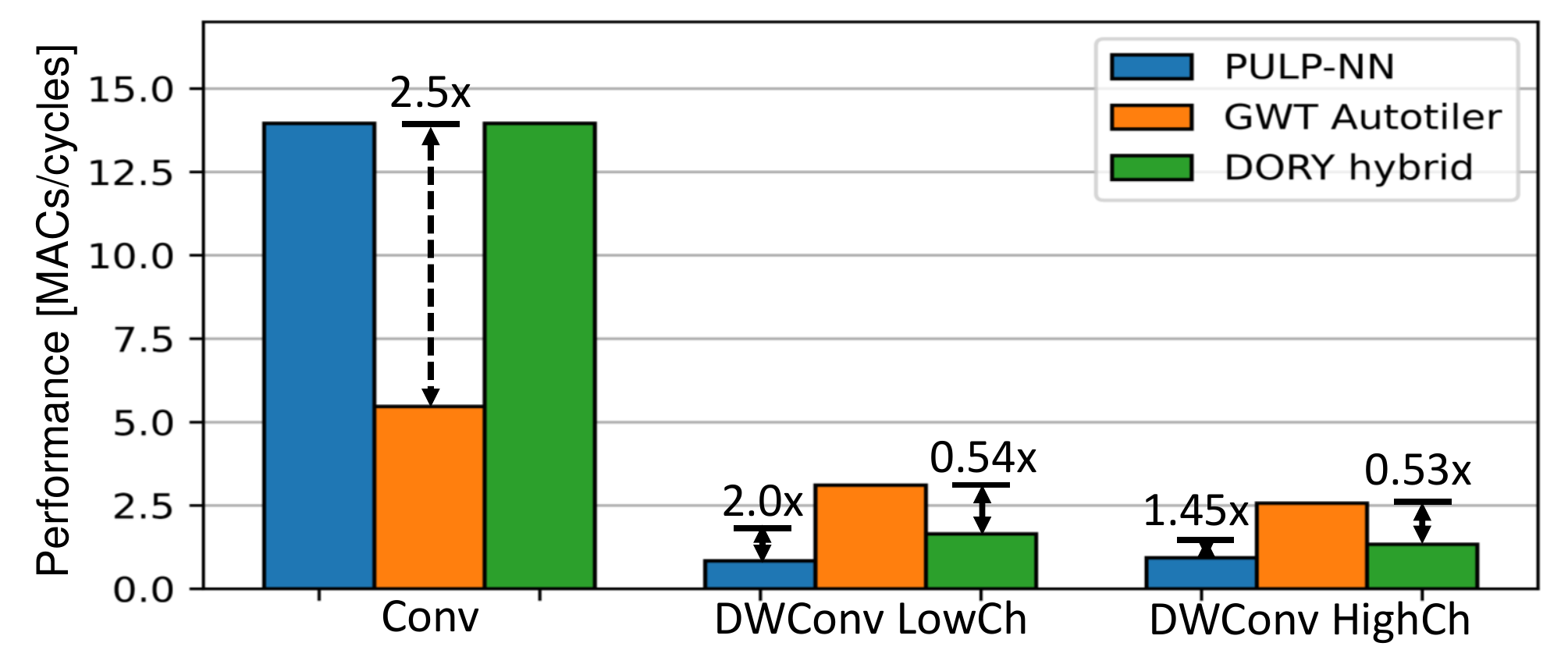}
\vspace{-0.5cm}
  \caption{Comparison between HWC, CHW, and DORY layers layout. Different kernels are explored.}
\label{fig:HWC_CHW}
\vspace{-0.4cm}
\end{figure}
Here, we discuss the improvement of the new DORY kernel library over PULP-NN kernels \cite{garofalo2020pulp} (HWC layout) and Greenwaves' ones (CHW layout).
In Fig. \ref{fig:HWC_CHW}, we show comparison on different layers, representative of the normal convolutions, and depth-wise ones.
On classical convolutions, our approach is $2.5\times$ faster compared to the CHW layout.
As discussed in Section \ref{sec:depthwise}, the DORY library includes an optimized depth-wise layer, reducing the penalty of using the HWC layout in its execution. Using an HWC layout on depth-wise layers can cause up to $3.7\times$ slow down if compared to the CHW one, strongly penalizing the performance for these layers.
We reduce this loss by a factor of 2: our kernel is $1.5\times$/$2.0\times$ faster than the HWC one, reaching $0.54\times$ the performance of the Greenwaves' one.
On the Mobilenet-v1-1.0 with resolution 128x128, updating the depth-wise and point-wise kernel from the HWC ones, we gain 1.79 MAC/cycles on the network's overall execution. 
At a frequency of 100~MHz on both cluster and I/O domains, we improved the 3.0 FPS of HWC layout, reaching 4.3 FPS thanks to the optimized DORY kernel library.

\revTCOMP{\subsection{Voltage and frequency scaling}}
%
\begin{figure}
  \centering
\includegraphics[width=0.85\columnwidth]{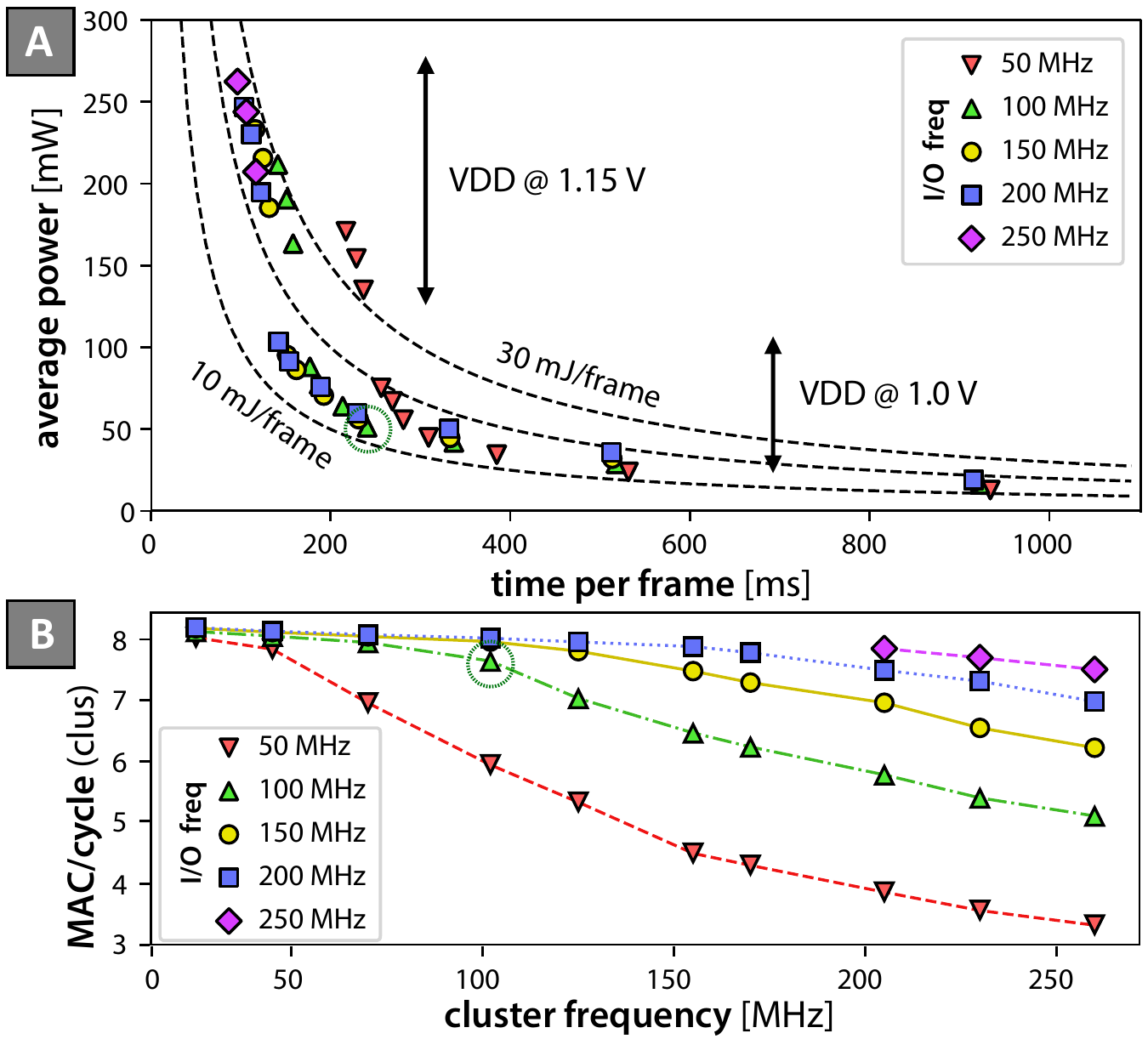}
\vspace{-0.3cm}
  \caption{Power, latency and MAC/cycles performance exploration with swiping frequencies. The 1.0-MobileNet-128 is used as a benchmark. CL frequency varies in [25 MHz, 260 MHz], I/O one in [50 MHz ,250 MHz]. A green dashed circle highlights the (100 MHz, 100 MHz) configuration that has been used throughout the paper.}
\label{fig:trade_off}
\vspace{-0.4cm}
\end{figure}

%
Since the I/O DMA and the cluster are in two different clock domains, the ratio of the two frequencies can significantly impact the bandwidth of both the $L3$-$L2$ and $L2$-$L1$ transfers and the performance and energy efficiency.
In Fig.~\ref{fig:trade_off}, we show the relationships between average power, execution time, and throughput in MAC/cycles, which are strictly related to the two frequencies.
Energy efficiency is also shown in sub-plot A as a set of iso-energetic curves.
A first significant effect that can be observed in these plots -- particularly sub-plot B -- is that increasing the fabric controller frequency strongly improves performance.
In fact, increasing the fabric controller frequency directly causes the L3-L2 memory transfers to be faster, minimizing the fraction of time in which the system is memory bound.
On the other hand, increasing frequencies also raises proportionally average dynamic power, as visible in sub-plot A.
However, the memory-boundedness increase is more detrimental to the overall energy efficiency, as can be observed for the case of the fabric controller running at \SI{50}{\mega\hertz}.
It is also interesting to observe that, using voltage and frequency scaling, it is possible to scale the execution of MobileNet from a minimum latency of \SI{93.9}{\milli\second} at \SI{24.6}{\milli\joule} per frame to minimum energy of 12.5 mJ at 244 ms per frame.

\revTCOMP{\subsection{Memory hierarchy sizing}}
\label{sec:mem_hier}
\begin{figure}
  \centering
\includegraphics[width=1.02\columnwidth]{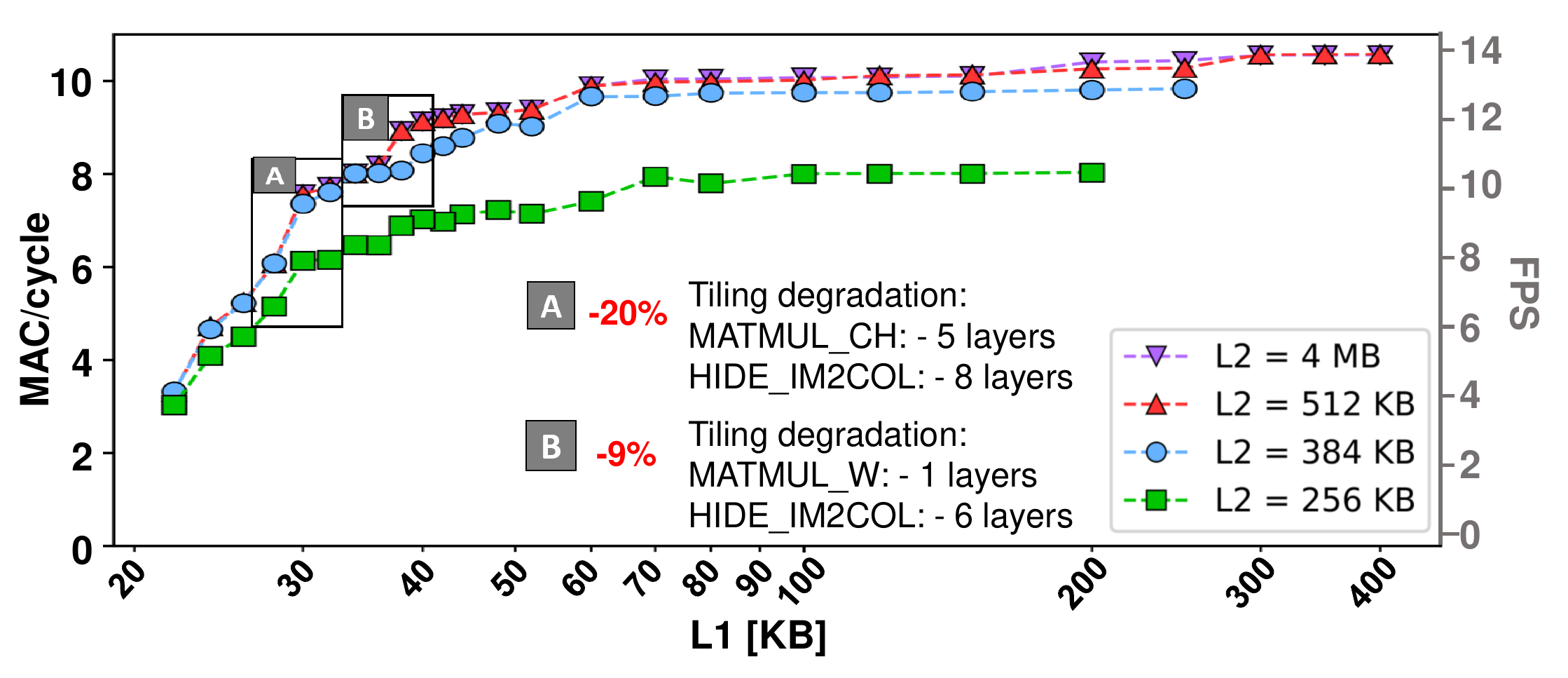}
\vspace{-0.5cm}
  \caption{MAC/cycles and FPS are explored with different configuration of L1-L2 memories using a 1.0-MobileNet-v1 with resolution 128x128. L2 varies from 256 kB (19/29 layers tiled from L3) to 4 MB (No L3 tiling), whereas L1 varies from 22 kB to 400 kB.}
\label{fig:trade_off_memory}
\vspace{-0.4cm}
\end{figure}
We also investigate the impact of memory dimensions on the network execution time.
To explore configurations with high dimensions of the memory, we used an FPGA-based emulator, realized with a Xilinx Zynq Ultrascale+ zcu102; the FPGA can host different instantiations of the PULP architecture template. 

%
Fig. \ref{fig:trade_off_memory} depicts MAC/cycles and FPS while sweeping L1 between [22 kB, 400 kB] and L2 in \{256 kB, 384 kB, 512 kB, 4 MB\}, highlighting different working corners in the tiling problem.
L1 memory limits have been chosen since \textit{i)} 22 kB are needed to construct the smaller tile available and store the corresponding im2col buffer, and \textit{ii)} over 400 kB no performance improvements are yet observed. 
L2 limits are related to chip design: while 256 kB is the lowest memory used as on-chip memory on a PULP platform~\cite{pullini2019mrwolf}, we foresee that 4 MB is the maximum memory that will be available in the near-future in low-cost IoT devices.
\revTCOMP{
The tool statically computes the minimum dimension of each memory level for the target DNN, raising an error if not compliant with input limits.
Then, it maximizes the occupation of the $Li$ buffers given as input constraints.
%
}

A first performance gap can be observed between the L2 = 256 kB and L2 = 512 kB configurations: with different L1 dimensions, using half of the memory causes up to 3.2 FPS loss @ 260MHz.
Using only half of the L2, 9 out of 29 layers demand the tiling of their activations from the external memory slowing down the execution of the first half of the network, since they can not fit the tightened constraint.
We can also observe a relatively constant decrease in performance when reducing L1 memory from 70 kB down to 22 kB with some abrupt performance loss.
Two different phenomena can be observed: \revTCOMP{\textit{i)} reducing L1 memory requires smaller tiles and hence more iterations, increasing overhead; \textit{ii)} reducing L1 memory too much can make the heuristics impossible to meet; for example, in case \textbf{A} of Fig. \ref{fig:trade_off_memory}, a reduction 30 kB to 28 kB causes this effect on 13 layers simultaneously, dropping performance by 20\%.
%
%
Conversely, from 70 kB to 400 kB of L1 the gain is minimal, because all the tiling heuristics are already satisfied.
}


Overall, thanks to DORY's optimized exploitation of memory bandwidth and locality enhancements due to backend and tiling, we see that a 80 kB L1 and 384 kB L2 memory configuration is sufficient to lead a MAC/cycle degradation of just 8\% (from 10.57 to 9.74 MAC/cycles) compared to the largest memory configuration for the targeted network (4 MB L2 and 400 kB L1, which eliminates external memory transfer and L2-L1 tiling) -- this results in a 91\%/80\% total L2/L1 memory size reduction in case this network is used to drive memory sizing.
%

\revTCOMP{
\subsection{Single core performance on different architectures}}
\label{sec:DORY_single_core}
\begin{figure}
  \centering
\includegraphics[width=0.98\columnwidth]{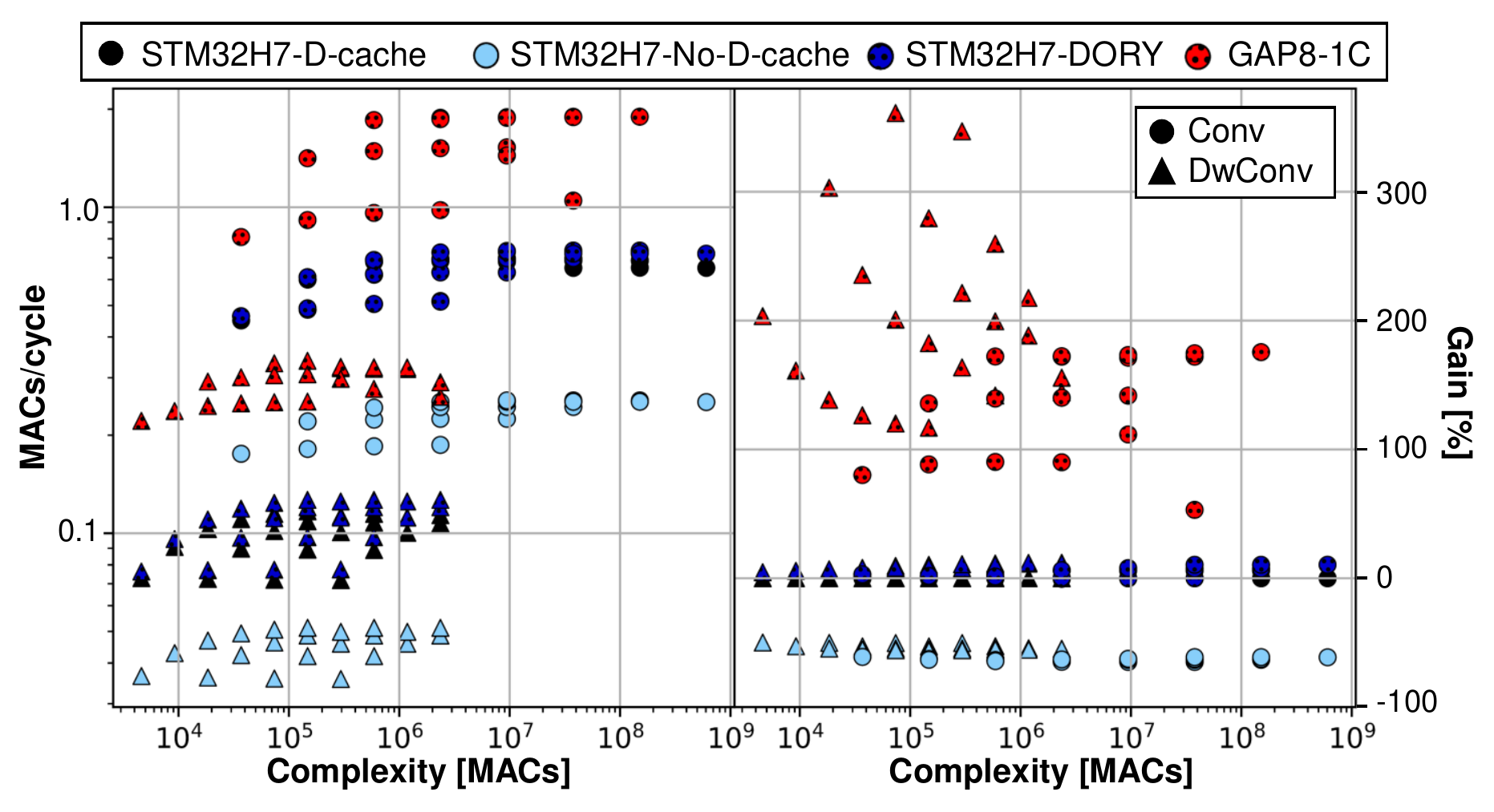}
\vspace{-0.3cm}
  \caption{\revTCOMP{On the left, absolute MAC/cycle of DORY framework on both STM32H7 and single-core GAP8, compared with default CUBE-AI/TensorFlow Lite for Micro layer backend, CMSIS-NN. On the right, relative gains compared to the fastest CMSIS-NN implementation. }}
\label{fig:SoA_1core}
\vspace{-0.4cm}
\end{figure}
%


\revTCOMP{
In this section, we explore the impact of architectural and microarchitectural choices on DNN deployment using DORY.
We do so by directly comparing the single-core performance obtained on GAP-8 with that achievable on a commercial STM32H743ZI2 MCU in several configurations.
This MCU features an ARM M7 core with 16 kB of D-Cache and a large two-banked 0-wait-states scratchpad of 128 kB called DTCM, allowing us to separately investigate the impact of software vs. hardware caching and that of the different microarchitectures.
%
}

\revTCOMP{
In our experiment, we tested 44 different configurations of layers (both depthwise and convolutional) spanning six orders of magnitudes of complexity.
We explored four sets of solutions: for GAP-8, we used DORY and run on a single core in the cluster; for the STM32H7, we used CMSIS-NN\footnote{We used CMSIS-NN instead of CUBE-AI due to its open-source nature; note that according to the ST forums, the fixed-point backend of CUBE-AI is ``based on the low-level ARM CMSIS-NN functions''.} with and without D-Cache enabled.
Finally, in the third STM32H7 configuration we ran using the DTCM scratchpad by combining DORY (for memory management) with CMSIS-NN.
This was possible thanks to the modular architecture of DORY and required only changing the computational backend and adapting the code generator to use the correct DMA hardware abstraction layer calls. }

\revTCOMP{
The results are shown in Fig.~\ref{fig:SoA_1core}.
First of all, as expected performance drops dramatically deactivating the D-Cache on the STM32: we observe a degradation of 58.5 $\pm$ 5.5 \% with respect to the baseline over all the benchmark layers.
%
More interestingly, our results also show that the software caching mechanism realized by DORY on the DTCM can achieve the same performance as the D-Cache on average, with a slight speedup in some cases: on average, 9.1$ \pm$ 2.1 \% for depthwise layers and 3.9 $\pm$ 3.8 \% for normal convolutions. }

\revTCOMP{
On the other hand, single-core execution on GAP-8 shows on average a speedup of 2.5$ \pm$0.9$\times$ with respect to the STM32H7 baseline in terms of cycle/cycle.
Since multi-core execution is disabled in this test, the speed up achieved in GAP8 with respect to the STM32H7 is referred mainly to the more specialized architecture, and in particular to the DSP extensions extensively exploited by the PULP-NN backend. }


\section{Conclusion}
\label{sec:conclusion}

In this work, we introduced a novel framework for DNN deployment, DORY, which unburdens the programmer from the manual optimizations of neural networks on end-nodes.
As a case study, we targeted a DNN-oriented MCU, GWT GAP-8, showing that it achieves 12.6$\times$ higher energy efficiency and 7.1$\times$ higher performance compared to the industry-standard STM32H743, and up to 26.6\% end-to-end inference improvement compared to the proprietary tool from GWT.
Our developments are released as open-source at {\texttt{\href{https://github.com/pulp-platform/dory}{https://github.com/pulp-platform/dory}}}.
Future work will focus on 
adding support for stronger quantization, hardware-accelerated primitives, and emerging memory technologies to support more high-accuracy networks directly on sub 10~mW extreme edge platforms.

\section*{Acknowledgement}
The authors thank Daniele and Margot Palossi for their help in setting up the RocketLogger to obtain GAP8 power traces.
%

\tiny


\begin{IEEEbiography}[{\includegraphics[width=0.82in,height=0.82in,clip,keepaspectratio]{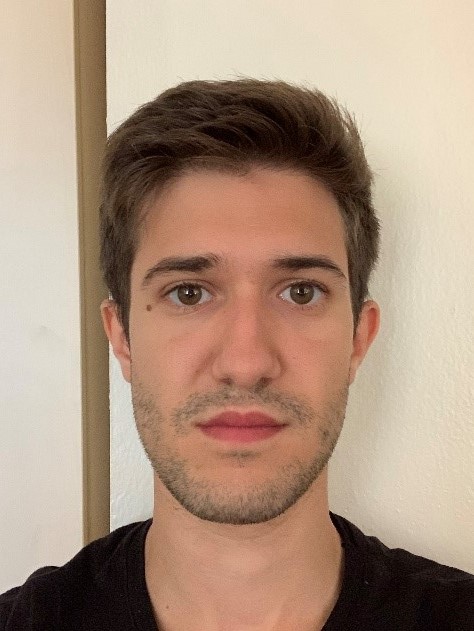}}]{Alessio Burrello}
received his B.Sc and M.Sc degree in Electronic Engineering at the Politecnico of Turin, Italy, in 2016 and 2018.  He is currently working toward his Ph.D. degree at 
University of Bologna, Italy.
His research interests include parallel programming models for embedded systems, machine and deep learning, hardware oriented deep learning, and code optimization for multi-core systems.
\end{IEEEbiography}

\begin{IEEEbiography}[{\includegraphics[width=0.82in,height=0.82in,clip,keepaspectratio]{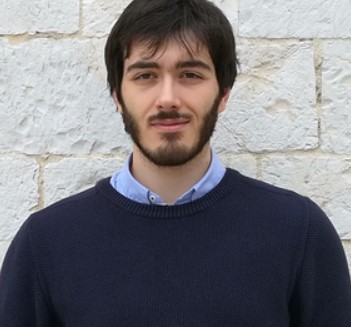}}]{Angelo Garofalo}
received the B.Sc and M.Sc. degree in electronic engineering from the University of Bologna, 
, Italy, in 2016 and 2018 respectively. He is currently working toward his Ph.D. degree at University of Bologna.
His main research topic is Hardware-Software design of ultra-low power multiprocessor systems on chip. His research interests include Quantized Neural Networks, Hardware efficient Machine Learning, and embedded architectures.
\end{IEEEbiography}

\begin{IEEEbiography}[{\includegraphics[width=0.82in,height=0.82in,clip,keepaspectratio]{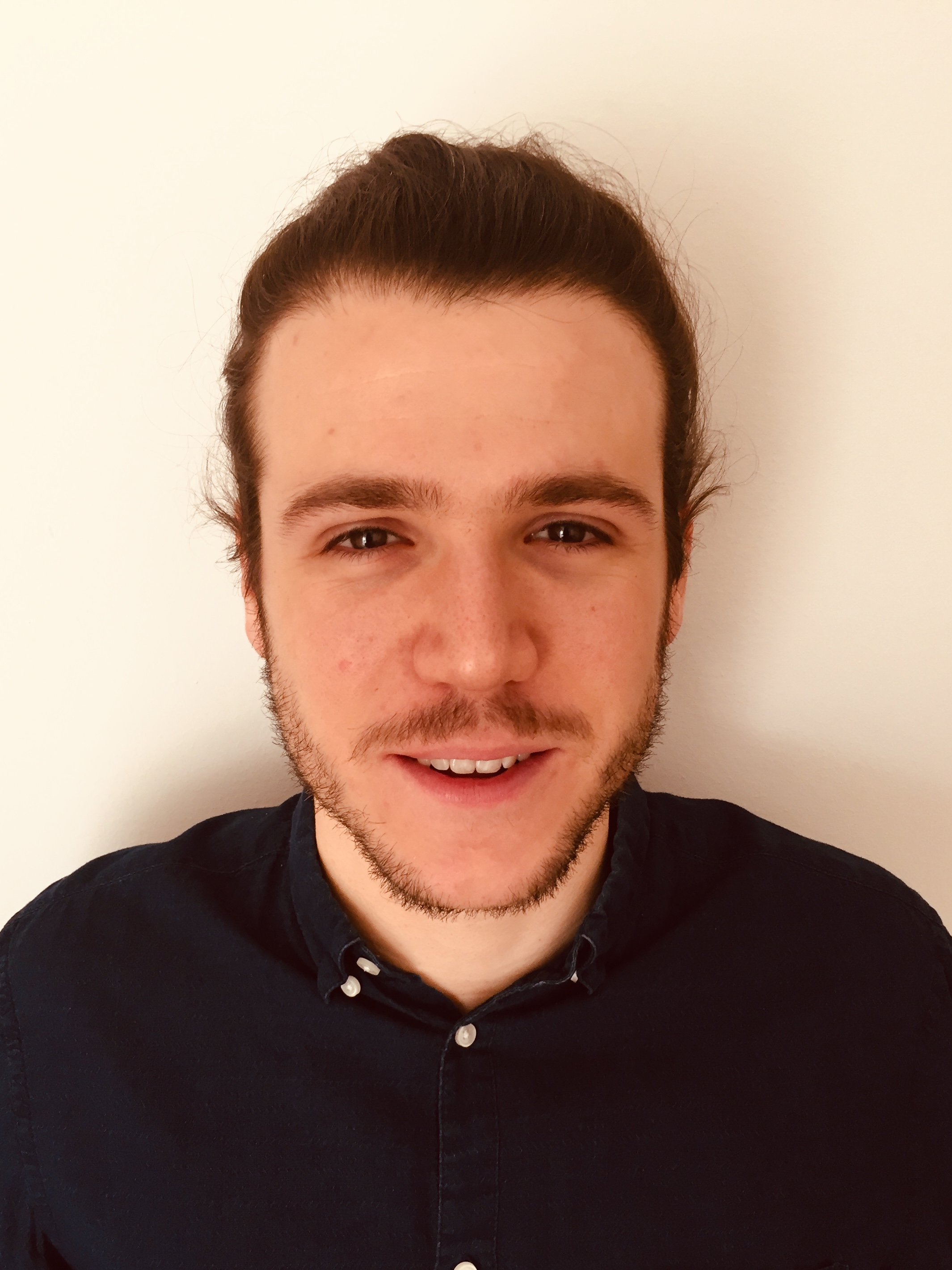}}]{Nazareno Bruschi} received the M.Sc degree in Electronic Engineering at the University of Bologna, Italy, in 2020. Since then, he is a Ph.D. student in the Department of Electrical, Electronic and Information Technologies Engineering (DEI) of the University of Bologna.
His research interests cover hardware and software optimization for low power and high efficiency embedded systems, parallel programming for multicore architectures and virtual prototyping.
\end{IEEEbiography}

\begin{IEEEbiography}[{\includegraphics[width=0.82in,height=0.82in,clip,keepaspectratio]{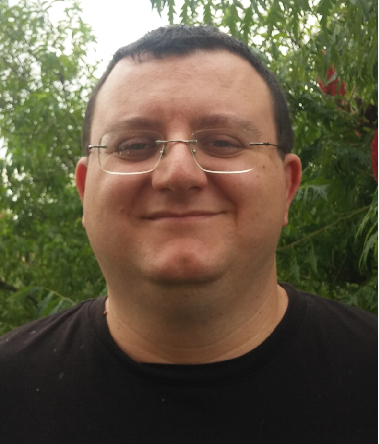}}]{Giuseppe Tagliavini}
received the Ph.D. degree in electronic engineering from the University of Bologna, Italy, in 2017.
He is currently an Assistant Professor at the University of Bologna. He has co-authored over 30 papers in international conferences and journals. His research interests include parallel programming models for embedded systems, and run-time optimization for multicore and many-core accelerators, and design of software stacks for emerging computing architectures.
\end{IEEEbiography}

\begin{IEEEbiography}[{\includegraphics[width=0.82in,height=0.82in,clip,keepaspectratio]{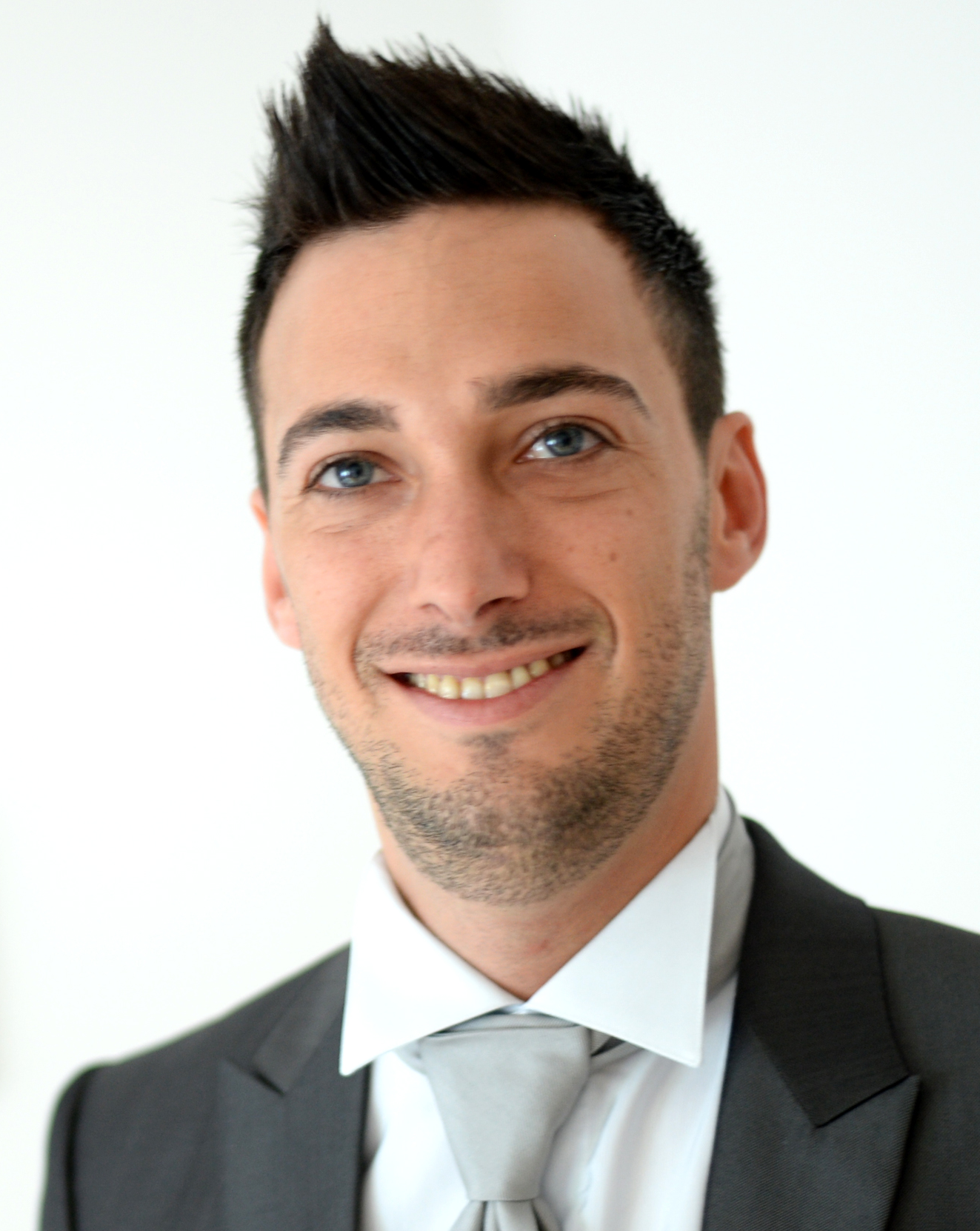}}]{Davide Rossi}, received the PhD from the University of Bologna, Italy, in 2012 where he currently holds an assistant professor position. His research interests focus on energy efficient digital architectures in the domain of heterogeneous and reconfigurable multi and many-core systems on a chip. 
This includes architectures, design implementation strategies, runtime support to address performance, and energy efficiency 
of ultra-low-power computing platforms. 
In these fields he has published more than 100 papers in international conferences and journals. 
He is recipient of Donald O. Pederson Best Paper Award 2018, - 2020 IEEE Transactions on Circuits and Systems Darlington Best Paper Award, 2020 IEEE Transactions on Very Large Scale Integration Systems Prize Paper Award.
\end{IEEEbiography}

\begin{IEEEbiography}[{\includegraphics[width=0.82in,height=0.82in,clip,keepaspectratio]{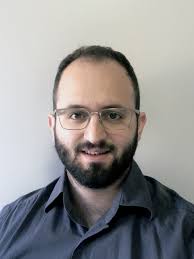}}]{Francesco Conti} received the Ph.D. degree in electronic engineering from the University of Bologna, Italy, in 2016. He is currently an Assistant Professor in the DEI Department of the University of Bologna. 
From 2016 to 2020, he held a research grant in the DEI department of University of Bologna and a position as postdoctoral researcher at the Integrated Systems Laboratory of ETH Zurich in the Digital Systems group.
His research focuses on the development of deep learning based intelligence on top of ultra-low power, ultra-energy efficient programmable Systems-on-Chip.
His research work has resulted in more than 40 publications in international conferences and journals
and has been awarded several times, including the 2020 IEEE TCAS-I Darlington Best Paper Award.
\end{IEEEbiography}

\end{document}